\documentclass[fleqn,10pt]{wlscirep}
\RequirePackage{doi}

\usepackage[utf8]{inputenc}
\usepackage{fullpage}
\usepackage{authblk}
\usepackage{amsmath,amsfonts,amssymb,amscd,latexsym}
\usepackage{caption,subcaption}
\usepackage[T1]{fontenc}
\usepackage[normalem]{ulem}
\usepackage{url}
\usepackage[bottom]{footmisc}
\usepackage{hyperref}
\hypersetup{
    colorlinks=true,
    linkcolor=blue,
    filecolor=magenta,
    urlcolor=cyan,
    citecolor=blue,
}

\usepackage{multicol,multirow}
\usepackage{makecell}
\usepackage{lscape}
\usepackage{booktabs}
\usepackage{colortbl}
\usepackage{longtable}
\usepackage{hhline}
\usepackage{hhline}
\usepackage{booktabs}
\usepackage{lineno}

\usepackage[version=3]{mhchem}

\newlength\savedwidth
\newcommand{\wcline}[1]{\noalign{\global\savedwidth\arrayrulewidth\global\arrayrulewidth 1.1pt} \cline{#1}
\noalign{\global\arrayrulewidth\savedwidth}}

\usepackage[ruled,vlined]{algorithm2e}

\title{
Shotgun crystal structure prediction using machine-learned formation energies
}

\author[1]{Chang Liu}
\author[2]{Hiromasa Tamaki}
\author[2]{Tomoyasu Yokoyama}
\author[2]{Kensuke Wakasugi}
\author[2]{Satoshi Yotsuhashi}
\author[1]{Minoru Kusaba}
\author[3]{Artem R. Oganov}
\author[1,4,*]{Ryo Yoshida}
\affil[1]{The Institute of Statistical Mathematics, Research Organization of Information and Systems, Tachikawa, Tokyo 190-8562, Japan}
\affil[2]{Technology Division, Panasonic Holdings Corporation, Kadoma, Osaka 571-8508, Japan}
\affil[3]{Skolkovo Institute of Science and Technology, Skolkovo Innovation Center, Moscow 121205, Russia}
\affil[4]{Department of Statistical Science, The Graduate University for Advanced Studies, Tachikawa, 190-8562, Japan}

\affil[*]{yoshidar@ism.ac.jp}


\begin{abstract}
Stable or metastable crystal structures of assembled atoms can be predicted by finding the global or local minima of the energy surface within a broad space of atomic configurations. Generally, this requires repeated first-principles energy calculations, which is often impractical for large crystalline systems. Here, we present significant progress toward solving the crystal structure prediction problem: we performed noniterative, single-shot screening using a large library of virtually created crystal structures with a machine-learning energy predictor. This shotgun method (ShotgunCSP) has two key technical components: transfer learning for accurate energy prediction of pre-relaxed crystalline states, and two generative models based on element substitution and symmetry-restricted structure generation to produce promising and diverse crystal structures. First-principles calculations were performed only to generate the training samples and to refine a few selected pre-relaxed crystal structures. The ShotunCSP method is computationally less intensive than conventional methods and exhibits exceptional prediction accuracy, reaching 93.3\% in benchmark tests with 90 different crystal structures.

\end{abstract}

\begin{document}

% \linenumbers

\flushbottom
\maketitle
\thispagestyle{empty}

% \noindent Please note: Abbreviations should be introduced at the first mention in the main text – no abbreviations lists. The suggested structure of the main text (not enforced) is as follows.

\section*{Introduction}

The prediction of stable or metastable crystal structures from a given chemical composition has remained a fundamentally unsolved task in solid-state physics for several decades \cite{martonak2003PredictingCrystal,oganov2006CrystalStructure}. In principle, the stable or metastable crystal structures of assembled atoms or molecules in the solid state can be determined using quantum mechanical calculations. Crystal structure prediction (CSP) is based on finding the global or local minima of the energy surface within a broad space of atomic configurations, in which the energy can be evaluated by first-principles density functional theory (DFT) calculations. The CSP problem can be solved by applying an exploratory algorithm to determine the crystal structure at the global or local minima by successively displacing the atomic configurations along the energy gradient.

A broad array of CSP methods have been developed to solve this problem, including brute-force random search \cite{pickard2006HighPressurePhases,pickard2007StructurePhase,pickard2011InitioRandom}, simulated annealing \cite{kirkpatrick1983OptimizationSimulated,pannetier1990PredictionCrystal}, the Wang--Landau method \cite{wang2001EfficientMultipleRange}, particle swarm optimization \cite{wang2010CrystalStructure,zhang2017ComputerAssistedInverse}, genetic algorithms \cite{oganov2006CrystalStructure,oganov2011HowEvolutionary,lyakhov2013NewDevelopments}, Bayesian optimization \cite{yamashita2018CrystalStructure}, and look ahead based on quadratic approximation (LAQA) \cite{terayama2018FinegrainedOptimization}. More recently, machine-learned interatomic potentials have attracted increasing attention because they can expedite the optimization process by bypassing time-consuming ab initio calculations \cite{jacobsen2018OntheFlyMachine,podryabinkin2019AcceleratingCrystal, takamoto2022UniversalNeural}. Conventionally, genetic manipulations such as mutation and crossover are performed to modify a set of candidate crystal structures, whereupon their DFT energies are used as goodness-of-fit scores to prioritize candidates for survival in the new generation. This process is repeated until the energy minima are reached. For example, the pioneering software USPEX implements a comprehensive set of genetic operations such as the mutation and crossover of crystal objects \cite{oganov2006CrystalStructure,lyakhov2013NewDevelopments,oganov2011HowEvolutionary}, while the CALYPSO code employs a genetic operation known as swarm shift \cite{wang2012CALYPSOMethod}. However, these algorithms are time-consuming because of the need for ab initio structural relaxation of the candidate crystals at every step of the optimization process. In response, CrySPY was developed to increase the computational efficiency by introducing a machine-learning energy calculator \cite{terayama2018FinegrainedOptimization} based on the Gaussian process regressor \cite{rasmussen2005GaussianProcesses}. The predictive performance is successively improved by accumulating a training set of candidate crystal structures and their relaxed energies via Bayesian optimization \cite{mockus1989BayesianApproach}. The surrogate energy predictor efficiently rules out unpromising candidates whose energies are unlikely to reach the minima. However, most existing methods utilize relaxed energy values to evaluate the goodness-of-fit in the selection process or to produce instances to train a surrogate model. This requires a large number of candidate structures to be relaxed at every step of the sequential search, which is impractical and computationally expensive for large systems that contain more than 30--40 atoms per unit cell. 

A promising solution is to fully replace ab initio energy calculations with machine-learning surrogates. Energy predictors trained using DFT property databases, such as the Materials Project \cite{jain2013CommentaryMaterials,MaterialsProject}, AFLOW \cite{curtarolo2012AFLOWAutomatic, Aflow}, OQMD \cite{kirklin2015OpenQuantum,oqmd}, and GNoME \cite{merchant2023ScalingDeep} databases, exhibit reasonably high prediction accuracy \cite{chen2019GraphNetworks, choudhary2021AtomisticLine,xie2018CrystalGraph}. However, models trained on instances of stable or metastable structures in such databases are unsuitable for the prediction of pre-relaxed energies for a given system \cite{gibson2022DataaugmentationGraph}. As shown later, although these models can predict energy differences between different crystalline systems, they cannot quantitatively discriminate between the energy differences of distinct conformations for the system of interest, which is a requirement for solving the CSP problem.

In this study, we employed a simple approach to building a predictive model for formation energies. First, a crystal-graph convolutional neural network (CGCNN) \cite{xie2018CrystalGraph} was trained using diverse crystals with stable or metastable states from the Materials Project database. Subsequently, for a given chemical composition, the energies of a few dozen randomly generated unrelaxed structures were calculated by performing single-point energy calculations, and a transfer learning technique \cite{weiss2016SurveyTransfer,yamada2019PredictingMaterials} was applied to fine-tune the pretrained CGCNN to the target system. Generally, limited data are available for model training, and randomly generated crystal conformations are distributed in high-energy regions. Models trained on such data are biased toward high-energy states and generally not applicable to the extrapolative domain of low-energy states in which optimal or suboptimal conformations exist. In CSP, a surrogate model must be able to predict the energies of various conformations, with high- to low-energy states corresponding to the pre- and post-relaxed crystal structures, respectively. We demonstrate that a surrogate model derived using transfer learning exhibits high prediction accuracy, even for low-energy states.

After creating candidate crystal structures, exhaustive virtual screening was performed using the transfer-learning-based energy predictor. The narrowed-down candidate crystals were relaxed by performing DFT calculations. A wide variety of structure generators can be applied to generate virtual crystal libraries, including (i) methods based on element substitution using existing crystal structures as templates \cite{hautier2011DataMined,wang2021PredictingStable,kusaba2022CrystalStructure}, (ii) atomic coordinate generators that consider crystallographic topology and symmetry \cite{bushlanov2019TopologybasedCrystal,fredericks2021PyXtalPython}, (iii) algorithms for reconstructing atomic configurations based on interatomic distance matrices (contact maps) predicted by machine learning \cite{hu2021ContactMap}, and (iv) deep generative models that mimic previously synthesized crystals \cite{kim2020InverseDesign,zhu2012ConstrainedEvolutionary,lee2021CrystalStructure}. In this study, we validated our framework using two sets of virtual libraries created using methods related to (i) and (ii), that is, element substitution of template crystal structures and a Wyckoff position generator for de novo CSP. The search space was narrowed down in the latter by machine-learning-based prediction of the space groups and Wyckoff-letter assignments. 

The ShotgunCSP workflow, which can be regarded as a high-throughput virtual screening of crystal structures, is perhaps the simplest among existing CSP methods to date. The entire workflow comprises first-principles single-point energy calculations for, at most, 3000 structures to create a training set for the transfer-learning-based energy predictor and the structural relaxation of a dozen or fewer narrowed-down candidate crystals in the final stage. Compared to conventional methods such as USPEX, the present method is significantly less computationally demanding. Furthermore, the prediction performance is outstanding, as confirmed by experimental validation; specifically, the method accurately predicted more than 90\% of the stable structures of 90 benchmark crystals with diverse space groups, structure types, constituent elements, system sizes, and application domains.

\begin{figure}[!htbp]
\centering
\includegraphics[width=\linewidth]{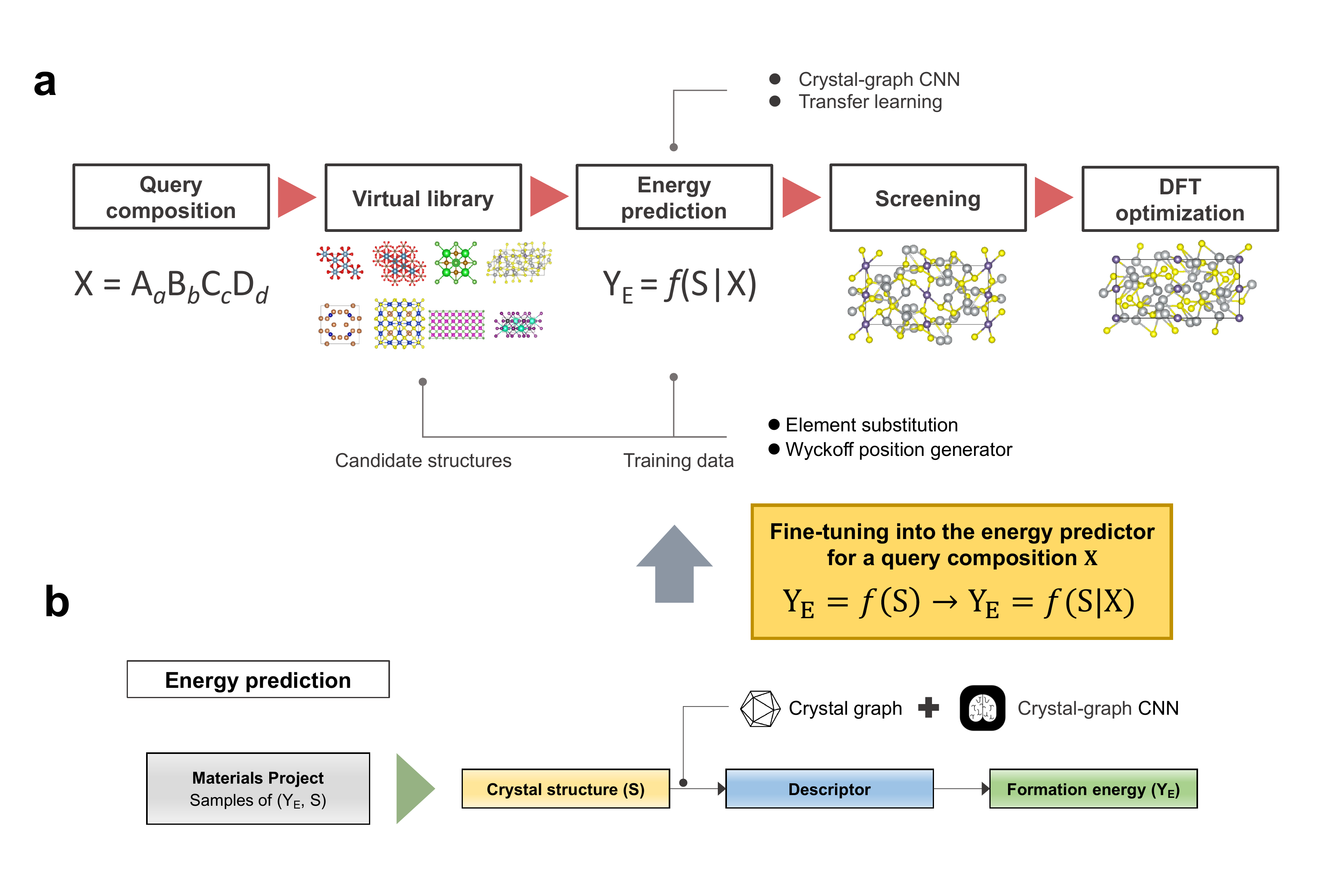}
\caption{Workflow of the shotgunCSP algorithm. (a) Virtual screening using a machine-learning-based energy predictor. A virtual library is created using a machine-learning-based element-substitution method (ShotgunCSP-GT) or a Wyckoff position generator (ShotgunCSP-GW). (b) Construction of formation-energy predictor based on a CGCNN. The CGCNN was trained with the Materials Project database and fine-tuned for the energy prediction of pre-relaxed candidate crystal structures for a query composition $X$.}
\label{fig:1}
\end{figure}

\begin{figure}[!htbp]
\centering
\includegraphics[width=\linewidth]{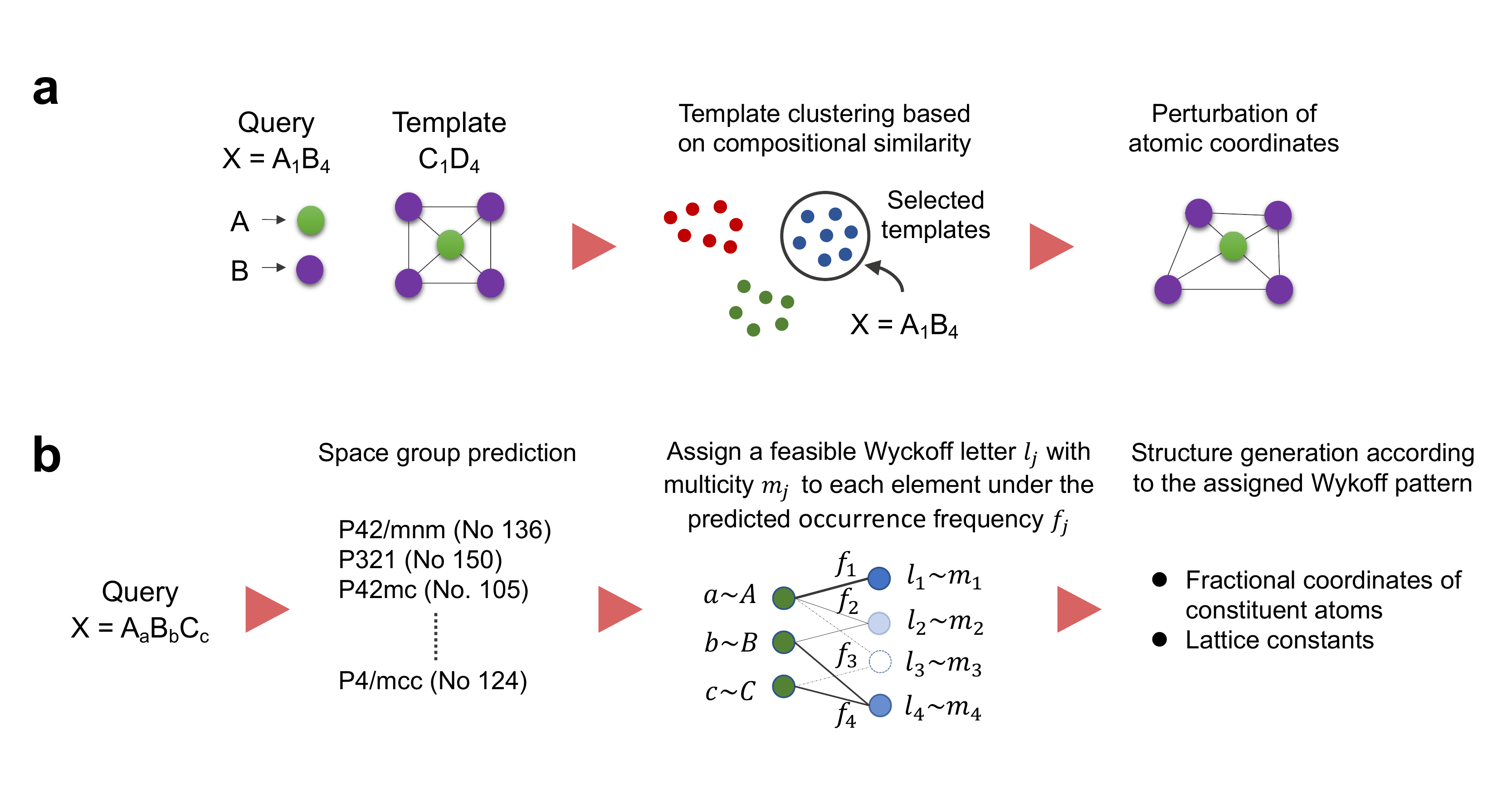}
\caption{Two different crystal structure generators are used to generate training instances to refine the energy prediction model and to create virtual libraries to be screened. (a) ShotgunCSP-GT: element substitution of template crystal structures in the Materials Project database. (b) ShotgunCSP-GW: Wyckoff position generator to produce symmetry-restricted atomic coordinates.}
\label{fig:generator}
\end{figure}

\section*{Results}

\subsection*{Outline of methods}

The stable crystal structures of atom assemblies with chemical composition $X$ were predicted using the machine-learning workflow summarized in Figure \ref{fig:1}. This method involves two key technical components: a high-performance surrogate model for predicting DFT formation energies and two different generative models for producing candidate crystal structures. 

For the energy calculation, a CGCNN with the same architecture as that in a previous paper \cite{xie2018CrystalGraph} was pretrained from scratch on a set of 126,210 crystals for which DFT formation energies are available in the Materials Project database. This model, referred to as the global model, can accurately predict the baseline formation energies of diverse crystal structures but is unable to discriminate between the local energy differences of different atomic conformations for a given target system. Therefore, the pretrained global model was localized to the target system $X$ by transfer learning. To this end, we randomly generated, at most, several thousand virtual crystal structures and calculated their formation energies by performing single-point energy calculations. The training structures were generated by either template-based substitution or the Wyckoff position generator, as described below. Using the dataset of generated structures, we performed transfer learning to adapt the pretrained global model to a local model applicable to the energy evaluation of different configurations for $X$. Here, the output layer was trained from scratch, whereas the pretrained weight parameters in the other layers were retained and fine-tuned (see Methods).

We developed and tested two algorithms to generate virtual crystals: element substitution and a Wyckoff position generator.
\begin{enumerate}
\item[]{{\bf Method 1 ---Element substitution (ShotgunCSP-GT)}:} Elements of already synthesized or theoretically possible crystals with the same composition as $X$ are randomly substituted (Figure \ref{fig:generator}(a)).

\item[]{{\bf Method 2 ---Wyckoff position generator (ShotgunCSP-GW)}:} For a target composition, with the space group given a priori or predicted, the generator randomly creates symmetry-restricted atomic coordinates from all possible combinations of Wyckoff positions (Figure \ref{fig:generator}(b)). A machine-learning predictor is employed to efficiently reduce the degrees of freedom in Wyckoff-letter assignment.
\end{enumerate}
The element-substitution method (ShotgunCSP-GT) cannot be applied unless a template is available for substitution, which limits its applicability. Therefore, we developed the Wyckoff position generator (ShotgunCSP-GW), which can generate novel structures when no template is available. The configuration of space groups and assignment of Wyckoff letters were deduced based on machine learning, as described below. The two generators were used to produce training instances for fine-tuning the CGCNN as well as candidate structures during high-throughput virtual screening. The workflows for the two generators differed slightly, as described below.

For ShotgunCSP-GT, crystal structures were generated by replacing elements in existing crystals. This mimics the process whereby humans synthesize new crystalline materials in a laboratory. For a given query composition $X$, we collected a set of template crystal structures with the same composition ratio as that of $X$ from the Materials Project database. Candidate structures were created by assigning the constituent elements of the query composition to the atomic coordinates of the selected templates. Elements with the same composition fraction in the template and query composition were substituted. When two or more elements had the same composition fraction, the assignment could not be uniquely determined. In this case, the most similar element pair was selected for substitution using the normalized Euclidean distance of the 58 element descriptors in the XenonPy library \cite{yamada2019PredictingMaterials,wu2020IQSPRXenonPy,liu2021MachineLearning,liu2023QuasicrystalsPredicted,xenonpy} as the similarity measure. The crystal structures with the substituted elements inherited the atomic coordinates of the template structures. The generated atomic coordinates were subjected to slight random perturbations as an additional refinement step.
Considering that multiple crystals in the database have the same prototype structure (for example, 8005 compounds have the composition ratio A$_1$B$_1$C$_2$), a cluster-based template selection procedure was introduced to select highly relevant templates with query composition $X$ while maintaining the diversity of the template structures. We applied DBSCAN \cite{ester1996DensitybasedAlgorithm,schubert2017DBSCANRevisited} to classify the templates into clusters in which the chemical compositions were converted into 290-dimensional compositional descriptors using XenonPy.
Then, only those templates belonging to the same cluster as the query composition $X$ were selected to identify a set of templates with high compositional similarity (see the Supplementary Information for a brief explanation of the DBSCAN algorithm). In addition, to eliminate structurally redundant templates, we used the StructureMatcher module of pymatgen \cite{ong2013PythonMaterials,StructureMatcher} to construct a unique set of templates without any identical prototype structures. The number of unique templates in the same cluster as the query composition is denoted by $K_{\mathrm{temp}}$. A virtual library was created by generating 1,000 structures from each of the $K_{\mathrm{temp}}$ selected templates by perturbing the atomic coordinates and lattice constants to form $1,000 \times K_{\mathrm{temp}}$ candidate structures. This procedure was also used to randomly generate ten structures for each template, which served as a training dataset of $10 \times K_{\mathrm{temp}}$ structures for the fine-tuning step. (See Methods for additional details.)

The ShotgunCSP-GW method produced random crystal structures with a prescribed space group for a given composition. The space group of the stable structure of a given composition $X$ was predicted based on machine learning. The assignment of Wyckoff letters to the constituent atoms was narrowed down using a predictive model trained on a given set of crystal structures in the Materials Project database, as described later. This predictive model enabled us to randomly generate promising Wyckoff patterns while eliminating wasted search space. By restricting the Wyckoff-site multiplicity and symmetry, the atomic coordinates and lattice parameters were generated uniformly from specific intervals. Structures generated with two or more atoms within a certain distance were excluded a posteriori. Here, a space-group predictor was used to estimate the space group of $X$ with the objective of predicting and limiting the space group of the stable crystalline state for a given composition $X$. We compiled a list of the chemical compositions and space groups of 33,040 stable crystal structures from the Materials Project database for the training set. Using this model, the space group of the crystal system for $X$ was restricted to the top $K_{\mathrm{SG}}$ candidates (where $K_{\mathrm{SG}}$ was set as 30). 
Based on this setting, $100 \times K_{\mathrm{SG}}$ training instances and $15,000 \times K_{\mathrm{SG}}$ candidate crystals were generated for the fine-tuning and virtual screening steps, respectively (See Methods for additional details).

The transfer learning-based energy predictor was then used for exhaustive virtual screening using each of the two generators separately. Finally, using DFT calculations, the promising structures that exhibited the lowest predicted energies were optimized with the Vienna Ab initio Simulation Package (VASP, version 6.1.2) \cite{kresse1996EfficientIterative} combined with projector augmented wave (PAW) pseudopotentials \cite{blochl1994ProjectorAugmentedwave} (see Methods for details). The top $K$ lowest-energy structures were subjected to structural relaxation with DFT. In this study, we selected $K=5 \times K_{\mathrm{temp}}$ and $K=10 \times  K_\mathrm{SG}$ structures for CSP using ShotgunCSP-GT and ShotgunCSP-GW, respectively. Generally, the top $K$ candidates with minimized energies had similar structures, many of which converged to the same crystal structure during structural relaxation. To eliminate this redundancy, we considered structural similarity when selecting the top $K$ candidate structures to maintain high structural diversity (see Methods).

\subsection*{Benchmark sets}
The performance of the proposed CSP algorithm was evaluated on three benchmark sets. The first  consisted of 40 stable crystals (Dataset I; Table \ref{tab:dataset_I_results}) that were selected based on a literature survey using two criteria: (i)  diversity of space groups, constituent elements, number of atoms, and element species; and (ii) diversity of applications such as battery and thermoelectric materials. Because the selection of crystal structures for Dataset I may have been biased due to the manual selection method, a second dataset of 50 stable crystals was randomly selected from the Materials Project database (Dataset II; Table \ref{tab:dataset_II_results}). For Datasets I and II, the numbers of atoms in the unit cells of the selected crystals were 2--104 ($\mathrm{mean}\pm\mathrm{standard\ deviation}$: $23.13\pm 24.09$) and 2--288 ($32.68\pm 45.41$), respectively (Figure S1). Of the benchmark crystals in Datasets I and II, 30\% had more than 30 atoms; owing to the computational complexity and search performance, solving these structures was expected to be difficult with conventional heuristic searches based on iterative first-principles calculations.

As a more challenging benchmark, we randomly selected 30 stable structures from the Materials Project database for which no template exists (Dataset III; Table S1). Most of the crystal structures in Dataset III had considerably more atoms per unit cell (22--152; $66.50 \pm 34.40$) than those in Datasets I and II.

\subsection*{Space-group prediction}
\begin{figure}[htbp]
\centering
\includegraphics[width=\linewidth]{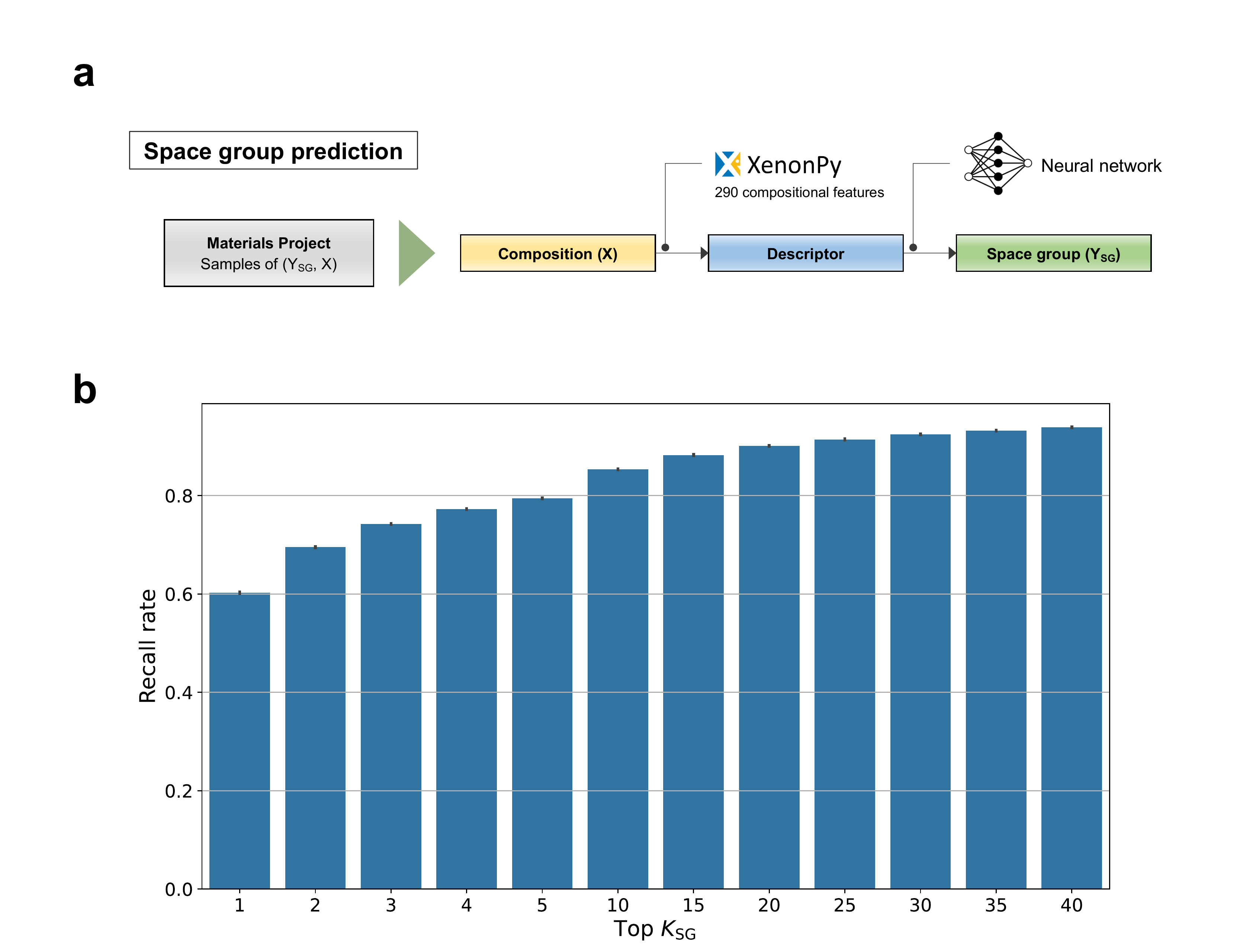}
\caption{Space-group prediction. (a) Machine-learning workflow. (b) Change in recall rate across the top 1 to the top 40 predictions of the space group. Narrowing down to the top 30 predicted labels should, on average, enable the inclusion of the true space groups for 92.61\% of the entire set of crystalline systems.}
\label{fig:recall}
\end{figure}

During virtual screening with the ShotgunCSP-GW, to narrow down the huge space of possible crystalline states, we introduced a multiclass discriminator to predict the space group $Y_{\mathrm{SG}}$ of a given chemical composition $X$ (Figure \ref{fig:recall}(a)). To train and test the classifier, we used 33,040 instances of chemical compositions with stable crystalline states, including 213 distinct space groups, compiled from the Materials Project database. The 17 remaining space groups were not registered in the database. The 120 benchmark crystal structures were removed from the training dataset. The compositional features of $X$ were encoded into the 290-dimensional descriptor vector using XenonPy \cite{yamada2019PredictingMaterials, wu2020IQSPRXenonPy, liu2021MachineLearning, xenonpy} (see Methods), and fully connected neural networks were trained to learn the mapping from the vectorized compositions to the 213 space groups. Of the total sample set, 80\% of the instances were used for training, and the remaining instances were used for testing. To statistically evaluate the prediction accuracy, training and testing were repeated independently 100 times. Details of the model construction, including hyperparameter adjustment, are provided in the Supplementary Information.

Figure \ref{fig:recall}(b) shows the change in the recall rate from the top 1 to the top 40 predictions; that is, the change in the proportion of true labels included in the top $K_{\mathrm{SG}}$ most probable predicted class labels ($K_{\mathrm{SG}} \in \{1,\ldots, 40\}$). The average recall rates in the top 1, 10, 30, and 40 predictions were $60.22 (\pm 0.87)\%$, $85.35 (\pm 0.54)\%$, $92.61 (\pm 0.49)\%$, and $94.02 (\pm 0.43)\%$, respectively. This indicates that focusing on the top 30 predicted labels allows us to identify the space groups for $92.61 (\pm 0.49)\%$ of the various crystalline systems. Using this model, the 213 space groups were narrowed down to the top 30 candidates, and for each of the selected candidates, a set of symmetry-restricted crystal structures was generated using ShotgunCSP-GW. 

To examine the differences in prediction performance by space group, we visualized the relationship between the top 30 recall rates and the training data size (Figure S2). The 25th, 50th, and 75th percentiles of the space-group-specific recall rates were 53.05\%, 72.62\%, and 84.25\%, respectively. Thus, the variability in recall rates was partially correlated with the number of training instances for each space group.

The distribution of composition ratios in the dataset was highly skewed, raising the concern that the prediction performance may vary greatly from one composition ratio to another. To address this concern, we conducted a robustness test by evaluating changes in the prediction accuracy upon varying the upper limit on the number of samples with the same composition ratio in the training set. The results were similar to those in Figure 3(b), demonstrating that the prediction performance did not vary significantly from one composition ratio to another (Figure~S3).

\subsection*{Wyckoff pattern prediction}

After narrowing down the space groups with machine learning, Wyckoff letters were randomly assigned to each atom to generate atomic coordinates. As the number of combinations of atoms and Wyckoff letters increases, the complexity of de novo CSP grows. For example, for the space group $lmma$ (No. 74) of $\mathrm{Mg_{8}B_{56}}$, the multiplicity of Wyckoff letters $\{a, b, c,d,e,f,g,h,i,j\}$ is $\{4,4,4,4,4,8,8,8,8,16\}$. In this case, the number of possible assignments for the Wyckoff letters exceeds 1755. If the assignments are incorrect, CSP usually fails. On the other hand, for the space group $la\bar{3}d$ (No. 230) of $\mathrm{Y_{24}Al_{40}O_{96}}$, the multiplicity of Wyckoff letters $\{a, b, c,d,e,f,g,h\}$ is $\{16,16,24,24,32,48,48,96\}$. Despite the substantial number of atoms in the unit cell, only 27 possible assignments exist. Successfully narrowing down the space groups and accurately predicting the assignments of Wyckoff letters is, therefore, expected to improve the CSP task significantly.

\begin{figure}[htbp]
\centering
\includegraphics[width=\linewidth]{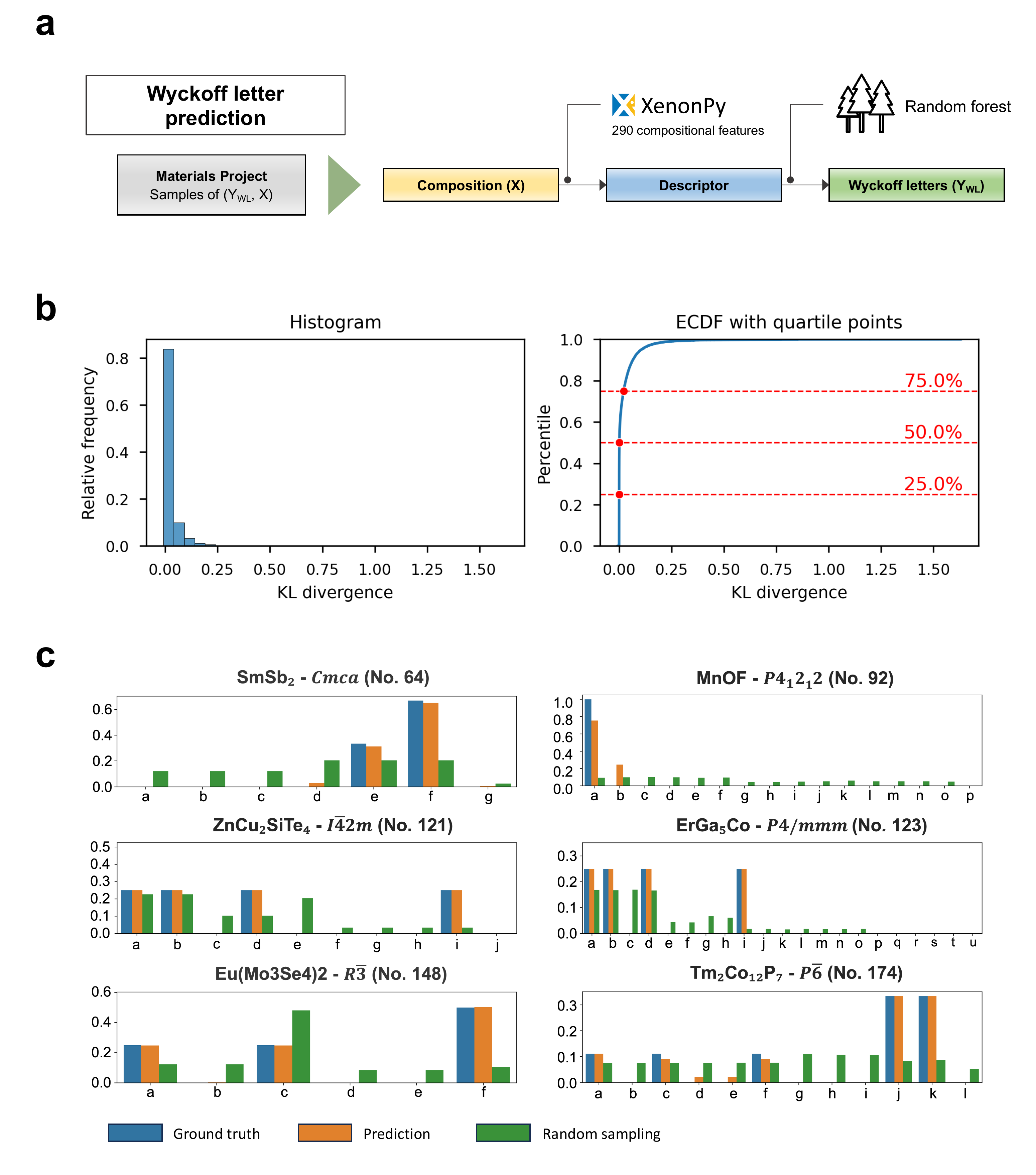}
\caption{Prediction of Wyckoff-letter assignments. (a) Machine-learning workflow. (b) Histogram and empirical cumulative distribution function (ECDF) of KL divergence between the relative occurrence frequencies and predicted probability distributions of Wyckoff letters for the test set. (c) Histograms of distributions of relative occurrence frequencies and predicted probabilities of Wyckoff letters for six randomly selected compounds, with their space-group information in parentheses.}
\label{fig:wp}
\end{figure}

This led us to construct a model to predict the occurrence frequencies of Wyckoff letters for stable structures based on the chemical composition (Figure \ref{fig:wp}(a)). A model was created for each space group, with the input chemical composition represented by a 290-dimensional descriptor using XenonPy. The output is the probability distribution of the occurrence of Wyckoff letters. Using the 33,040 instances of stable structures in the Materials Project database (excluding the 120 benchmark crystals; 80\% were used for training and 20\% for testing), we trained a random forest regressor for each space group.

Figure \ref{fig:wp}(b) summarizes the prediction accuracy for the test set. The discrepancies between the output probability distribution $p_1, \ldots, p_M$ of the trained model and the actual relative frequencies $q_1, \ldots, q_M$ of $M$ Wyckoff letters were measured based on the Kullback--Leibler (KL) divergence:
\begin{eqnarray}
\sum_{i=1}^M q_i \log \frac{q_i}{ p_i }
\end{eqnarray}
The distribution of KL divergence for the test set was found to be highly concentrated around zero in most cases (Figure \ref{fig:wp}(b)). This indicates that the Wyckoff letters for stable structures are predictable from the chemical composition.

Using the Wyckoff-letter occurrence probability for a query composition, we randomly assigned Wyckoff letters using the procedure in the Methods section. The possible assignment of Wyckoff letters to elements is constrained by their multiplicity and query composition ratio. The sampling algorithm was designed to satisfy these constraints;  the frequency of generated Wykoff labels generally matches the probabilities predicted by the random forest regressors.

Figure \ref{fig:wp}(c) shows the differences in the Wyckoff patterns generated with and without the Wyckoff-letter predictor by comparing the frequencies of actual and sampled Wyckoff letters for six randomly selected crystals. In all cases, the frequencies of Wyckoff letters generated from the predictor agreed well with the true frequencies. By contrast, those of randomly generated Wyckoff letters deviated significantly from the true frequencies.

\subsection*{Energy prediction}
\begin{figure}[htbp]
\centering
\includegraphics[width=\linewidth]{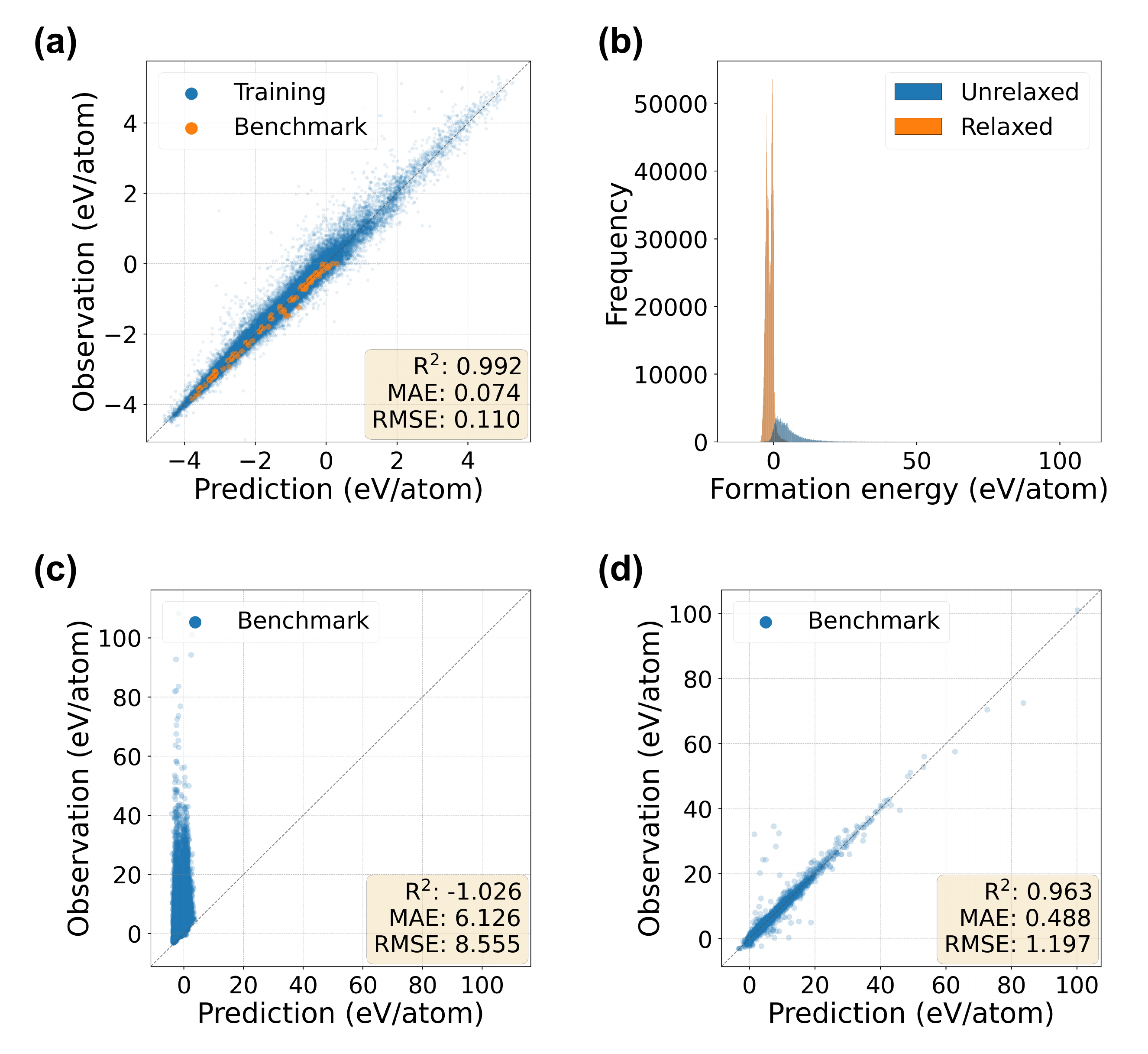}
\caption{Performance of CGCNN for the prediction of DFT formation energies with and without transfer learning. The root mean square error (RMSE), mean absolute error (MAE), and coefficient of determination ($\text{R}^2$) for test instances are shown on each parity plot. (a) Results of global model for the prediction of relaxed formation energies of 90 benchmark crystals (orange). (b) Histogram of DFT formation energies of relaxed and randomly generated pre-relaxed structures. (c, d) Prediction of pre-relaxed formation energies of 100 randomly generated conformations for each of the 90 benchmark systems (c) without and (d) with fine-tuning of the pretrained global energy prediction model.}
\label{fig:3}
\end{figure}

The global energy prediction model was constructed by training the CGCNN on 126,210 stable and metastable crystal structures whose formation energies were retrieved from the Materials Project database; the 120 benchmark crystals were excluded from the training set. To validate the predictive capability and uncertainty of the global model, we randomly extracted 80\% of the overall dataset and created 100 bootstrap sets. The average mean absolute error (MAE) across the 25,249 test cases reached 0.074 eV/atom, with a standard deviation of 0.003, which is comparable to that in previous studies \cite{xie2018CrystalGraph}. Figure \ref{fig:3}(a) shows the prediction results for the 90 benchmark crystals in Datasets I and II.

Note that the global model is unsuitable for predicting the energies of different randomly generated conformations for each composition $X$ (Figure \ref{fig:3}(c)). In addition, it failed to discriminate between the energies of different randomly generated conformations for the 90 benchmark crystals. We tested the predictive capability of the global model on the DFT energies of 100 randomly generated pre-relaxed conformations for each of the 90 benchmark crystals. The average MAE decreased to 6.126 eV/atom with a standard deviation of 2.010. A similar result was obtained when the global model was trained with approximately 1,021,917 instances of OQMD, including the formation energies of both relaxed and unrelaxed structures.

To overcome this limited predictive ability, the pretrained global model was localized to the target system $X$ by transfer learning. For each $X$, the formation energies of a maximum of 3000 virtual crystals, generated as described above, were obtained by DFT single-point energy calculations, and the pretrained global model was fine-tuned to the target system. As shown in Figure \ref{fig:3}(d), transfer learning improved the prediction performance for the formation energies of the 9000 additional conformations. The average MAE reached 0.488 eV/atom with a standard deviation of 0.453 eV/atom. This is a 12.6-fold improvement compared to the that of the pretrained global energy prediction model.

\subsection*{CSP using the library generator based on element substitution}
We used the fine-tuned surrogate energy predictor to sort the virtual crystals generated by the element-substitution method, narrowed them down to the top five structures for each template as described above, and then performed structural relaxation using DFT. The $J$ relaxed structures with the lowest DFT energies were used as the final set of predicted structures. Figure \ref{fig:4} shows the top two ($J=2$) predictions and the true structures for some selected cases. Figure S6 shows the top two predicted structures for all 120 benchmark crystals. Tables \ref{tab:dataset_I_results} and \ref{tab:dataset_II_results} summarize the success or failure of the top $5 \times K_\mathrm{temp}$ predictions for all crystal systems in Datasets I and II. The accuracies for Datasets I and II were 82.5\% and 86.0\%, respectively. No significant differences were observed in the accuracies between the two benchmark sets. Interestingly, increasing the number of atoms in the unit cell did not degrade the performance. Among the failure cases, five instances (\ce{NaCaAlPHO5F2}, \ce{K20Ag8As12Se36}, \ce{Na1W9O27}, \ce{Na80Fe16P32O128F32}, and \ce{Y8Si10Ir18}) did not have template structures in the Materials Project database with identical composition ratios.  
Excluding the five cases without a template, the accuracies for Datasets I and II reached 84.6\% and 93.5\%, respectively. For the other nine failure cases, for which a template was available but prediction failed, none of the Wyckoff-letter patterns of the true structures were included in the template structure set in the Materials Project database.

In summary, crystal systems with one or more replaceable template structure can be largely predicted by substituting the elements in existing crystals. For example, in the Materials Project database, the proportion of crystals with one or more interchangeable template structures was 98.0\%. A similar conclusion was reached with an element-substitution-based CSP machine-learning algorithm called CSPML \cite{kusaba2022CrystalStructure}. The prediction accuracy of the current element-substitution-based shotgunCSP for Datasets I and II was higher than that of CSPML, which was 65.6\% for the top ten predictions (see Table S4 for the testing results and the Supplementary Information for an outline of the methods).

\subsection*{CSP using the Wyckoff position generator}

\begin{figure}
\centering
\includegraphics[width=0.95\linewidth]{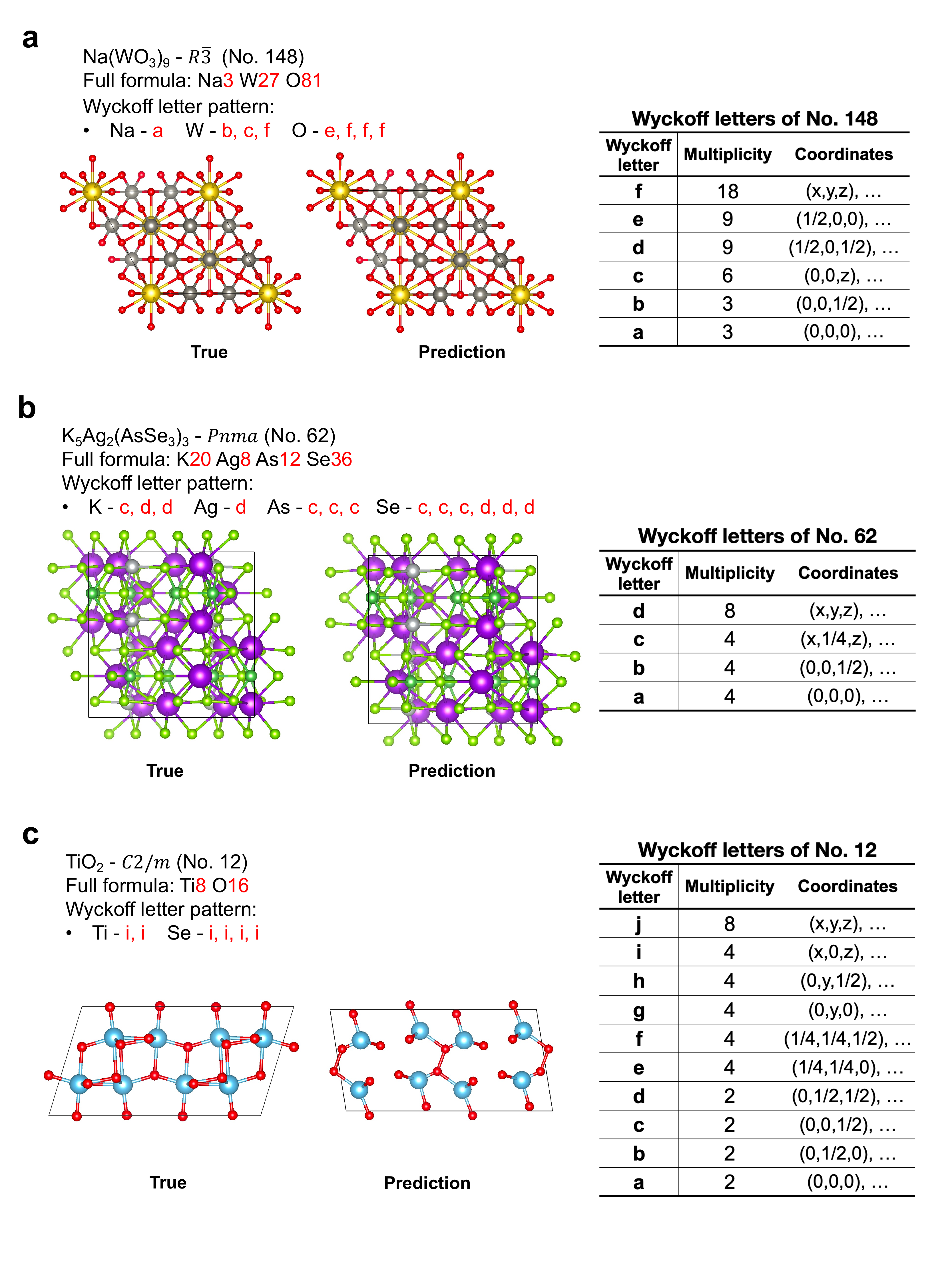}
\caption{Three crystal structures predicted with ShotgunCSP-GW: \ce{Na(WO3)9} (full formula: \ce{Na3W27O81}, space group: $R\bar3$), \ce{K5Ag2(AsSe3)3} (full formula: \ce{K20Ag8As12Se36}, space group: $Pnma$), and \ce{TiO2} (full formula: \ce{Ti8O16}, space group: $C2/m$). For \ce{Na(WO3)9}, the number of possible Wyckoff-letter combinations is limited to 48 when the space group is considered. For \ce{K5Ag2(AsSe3)3}, despite the number of possible Wyckoff-letter combinations exceeding 300, the Wyckoff-letter assignment predictor reduces the search space considerably and effectively. For \ce{Ti8O16}, which the CSP algorithm failed to predict, the high degree of freedom in the coordinate configurations prevented the generation of promising atomic coordinates despite the successful Wyckoff-letter assignment.}
\label{fig:complicated_structures}
\end{figure}

The top ten candidate structures predicted using the Wyckoff position generator with the lowest surrogate energies for each predicted space group were selected for structural relaxation using DFT ($J = 10 \times K_{\mathrm{SG}}$ with $K_{\mathrm{SG}}=30$). Figure \ref{fig:4} displays the top two predicted and true structures for some selected examples, and Figure S6 shows the top two predicted structures for all 120 benchmark crystals. For the top ten predicted structures, 77.5\% and 78.0\% of the known stable structures were accurately predicted for Datasets I and II, respectively. Tables \ref{tab:dataset_I_results} and \ref{tab:dataset_II_results} summarize the performance (success or failure) of the top ten predictions for all crystal systems in Datasets I and II, respectively. The overall performance was lower than that of the element-substitution-based CSP algorithm. One reason for the decreased accuracy is the failure of space-group prediction. Specifically, the model failed to predict the space groups for approximately 5\% of the 90 benchmark crystal structures in Datasets I and II. This is almost the same level of accuracy as reported above.

shotgunCSP using the Wyckoff position generator successfully predicted 31 and 38 of the crystals in Datasets I and II, respectively. To highlight the prediction mechanism of the proposed method, we focus here on three crystals that were successfully predicted but for which no template exists in the Materials Project database: \ce{Y4Si5Ir9}, \ce{K5Ag2(AsSe3)3}, and \ce{Na(WO3)9}. These compounds all have relatively large numbers of atoms in the unit cells (36, 76, and 111, respectively). Nevertheless, for the space group $R\bar3$ (No. 148) of the stable structure of \ce{Na(WO3)9} (Figure \ref{fig:complicated_structures}(a)), because the Wyckoff letters $\{a, b, d, e\}$ are coordinate-fixed, the number of possible combinations of Wyckoff letters is reduced to approximately 48 owing to its multiplicity constraints. Consequently, the effective dimensions of the search space could be reduced by considering the crystal symmetry. This explains why shotgunCSP successfully predicted the complex stable structure of \ce{Na(WO3)9}. For \ce{K5Ag2(AsSe3)3}, which has 76 atoms in its unit cell (space group $Pnma$; No. 62) (Figure~\ref{fig:complicated_structures}(b)), the possibility of replacing the Wyckoff letter $\{c\}$ with $\{a\}$ or $\{b\}$ increases the number of possible Wyckoff-letter combinations to over 300, even when considering multiplicity constraints. Nevertheless, the shotgunCSP achieved success, primarily because the occurrence probability of Wyckoff letters $\{a\}$ and $\{b\}$ was successfully predicted to be extremely low by the Wyckoff-letter assignment predictor. This significantly narrowed the extensive search space during candidate structure generation, highlighting the significance of the Wyckoff-letter assignment refinement strategy in the success of the CSP task. The same mechanism facilitated the prediction of \ce{Y4Si5Ir9}.

On the other hand, of the 85 crystals for which the space groups were correctly identified, the Wyckoff-position-generator-based CSP could not predict 9 and 11 of the true stable structures of crystals in Datasets I and II, respectively. Of these failure cases, true Wyckoff pattern generation failed in 3 and 3 cases for Datasets I and II, respectively, while ground-truth generation failed in 6 and 8 cases, respectively, despite correct true Wyckoff pattern generation. To elucidate the origin of these failures, we examined the generated structures in detail. The majority of structures that could not be predicted were characterized by low-symmetry structures with a space-group number below 142, particularly below 15, such as orthorhombic, monoclinic, and triclinic structures. Because of their low symmetry, structures belonging to these space groups have high degrees of freedom in their coordinate configurations. Furthermore, the number of combinations of Wyckoff patterns with the same multiplicity is greater for lower-symmetry space groups. For instance, space group $C2/m$ (No. 12) has one coordinate-free Wyckoff letter $\{j\}$ with multiplicity 8; three coordinate-free Wyckoff letters $\{g, h, i\}$ and two coordinate-fixed Wyckoff letters $\{e, f\}$ with multiplicity 4; and four coordinate-fixed Wyckoff letters $\{a, b, c, d\}$ with multiplicity 2. Their possible combinations form an extensive search space. This was the case for \ce{TiO2} (full formula: \ce{Ti8O16}); despite the successful prediction of the Wyckoff-letter configuration ${\mathrm{Ti}: 4i, \mathrm{Ti}: 4i, \mathrm{O}: 4i, \mathrm{O}: 4i, \mathrm{O}: 4i, \mathrm{O}: 4i}$, the generation of precise atomic coordinates was unsuccessful (Figure~\ref{fig:complicated_structures}(c)).

Many of the predicted structures that were determined to be failures have been experimentally reported as metastable structures (Figure \ref{fig:4}).  For example, for \ce{TiO2} \cite{cui2016FirstprinciplesStudy}, the anatase type is the most stable structure according to first-principles calculations, whereas the predicted structure was the rutile type, which is a known metastable state of \ce{TiO2}. Similarly, the true stable structure of \ce{Si3N4} \cite{kroll2003PathwaysMetastable} is the hexagonal structure, $\beta$-\ce{Si3N4}, whereas the predicted structure was the willemite-II type, which has been identified as a metastable structure of \ce{Si3N4} by DFT calculations. In many cases, even the predicted crystal structures that were judged to be failures partially captured structural features that were similar to those of the true structures. For example, the predicted structures of \ce{ZrO2} and \ce{LiP(HO2)2} did not exactly match the true structures, but differed only slightly in atomic positions (Figure \ref{fig:4}). The energy difference between the true and predicted structures of these compounds was <5 meV/atom. Although the prediction of stable structures for low-symmetry compounds was often not entirely accurate, metastable structures or partial structural patterns can be predicted using this method.

For the cases where the prediction failed, the energy differences between the predicted and ground-truth structures are briefly discussed in the Supplementary Information. Although there were a few cases where the energy of the predicted structure was lower or almost the same as that of the ground-truth structure, we decided not to delve deeper into these cases.

We also evaluated the predictive performance of our method in quite challenging scenarios using Dataset III, which resulted in a notably low accuracy of 6.7\% (see Table S1). Despite this outcome, the rather complex crystal structures of \ce{Al8(Pb3O7)3} and \ce{Mg10B16Ir19} were accurately predicted. However, predicting crystal structures for more complex systems would require additional computational resources and enhanced methodologies.

\subsection*{Comparison with USPEX}

The CSP tasks for Datasets I and II were conducted using USPEX by applying the calculation conditions outlined in the Methods section, with the assumption that the ground-truth space groups are known. The USPEX calculations were executed on the SQUID supercomputer system at Osaka University, which has two Intel Xeon Platinum 8368 CPUs with 76 cores running at 2.40 GHz at each node \cite{SQUID}. The calculation of each crystal structure was allocated to one node, with the number of MPI cores set to 38 when the number of calculated atoms was less than 38, and to 76 otherwise.

Using the settings described in the Supplementary Information, only tasks involving small-unit-cell systems (comprising approximately 20 or fewer atoms in the primitive unit cell) were accomplished successfully. Specifically, within the allocated computational resources, 13 and 12 systems were completed for Datasets I and II, respectively. For the completed tasks, USPEX had prediction accuracies of 92.3\% and 91.7\% for the benchmark crystals in Datasets I and II, respectively (Table S5). The median number of structure relaxation calculations performed was 167, with computations taking 37.7 h on the designated supercomputer system. In comparison, our method utilizing the with ShotgunCSP-GW generator for the same set of 25 benchmark structures yielded accuracies of 84.6\% and 83.3\% for Datasets I and II, respectively (Table S5). The median number of structure relaxation calculations performed was 172, with computations taking 21.4 h. While USPEX was conducted with the true space  as an initial parameter (where other space groups were also searched during the random search process), our method searched for the top 30 space groups. Thus, our noniterative approach, utilizing an energy predictor to identify structures with sufficiently low energy followed by structure relaxation, reached the same level of accuracy as USPEX while significantly improving the computational speed.

\renewcommand\theadalign{cc}
\renewcommand\theadfont{\bfseries}
\renewcommand\theadgape{\Gape[3pt]}
\renewcommand\cellgape{\Gape[4pt]}

% {\setlength{\tabcolsep}{4pt}
\begin{longtable}[!htbp]{r|cccccc}
\caption{
Results of the CSP algorithm with the ShotgunCSP-GT and ShotgunCSP-GW generators for the 40 crystals comprising Dataset I. The top $5 \times K_{\mathrm{temp}}$ or top $10 \times K_{\mathrm{SG}}$ (where $K_{\mathrm{SG}} = 30$) virtual structures with the lowest surrogate energies from element substitution and the Wyckoff position generator, respectively, were selected for DFT structural relaxation. The second column indicates the number of atoms in the primitive unit cells. In the fourth and fifth columns, symbols $\checkmark$ and $\times$ denote prediction success and failure, respectively, and dashes ($-$) indicate cases for which there was no template for element substitution or for which the calculation failed. For the fourth column, the left and right sides of the parentheses indicate the success or failure of Wyckoff-letter generation and space-group prediction, respectively. For the fifth column, the left and right sides of the parentheses indicate whether a sufficiently similar template structure ($\tau \leq 0.2$) was found and  the total number of templates selected ($K_{\mathrm{temp}}$), respectively.
 }

\label{tab:dataset_I_results}\\
\wcline{1-5}
    \thead{Composition} & \thead{Number of atoms} & \thead{Space group} & \thead{ShotgunCSP-GW} & \thead{ShotgunCSP-GT} \\
\cline{1-5}
\endfirsthead 
    
\multicolumn{5}{l}{Table \ref{tab:dataset_I_results} continued}\\
\wcline{1-5}
    \thead{Composition} & \thead{Number of atoms} & \thead{Space group} & \thead{ShotgunCSP-GW} & \thead{ShotgunCSP-GT} \\
\cline{1-5}
\endhead
    
\wcline{1-5} 
\endfoot
    \endlastfoot
        \ce{C}              & 4 & $R\bar3m$      & $\checkmark$ ($\checkmark$ \slash\ $\checkmark$) & $\checkmark$ ($\checkmark$ \slash\ 37) \\ \hline
        \ce{Si}             & 2 & $Fd\bar3m$     & $\checkmark$ ($\checkmark$ \slash\ $\checkmark$) & $\checkmark$ ($\checkmark$ \slash\ 16) \\ \hline
        \ce{GaAs}           & 2 & $F\bar43m$     & $\checkmark$ ($\checkmark$ \slash\ $\checkmark$) & $\checkmark$ ($\checkmark$ \slash\ 27) \\ \hline
        \ce{ZnO}            & 4 & $P6_3mc$       & $\checkmark$ ($\checkmark$ \slash\ $\checkmark$) & $\checkmark$ ($\checkmark$ \slash\ 141) \\ \hline
        \ce{BN}             & 4 & $P6_3/mmc$     & $\checkmark$ ($\checkmark$ \slash\ $\checkmark$) & $\checkmark$ ($\checkmark$ \slash\ 137) \\ \hline
        \ce{LiCoO2}         & 16 & $R\bar3m$     & $\checkmark$ ($\checkmark$ \slash\ $\checkmark$) & $\checkmark$ ($\checkmark$ \slash\ 136) \\ \hline
        \ce{Bi2Te3}         & 5 & $R\bar3m$      & $\checkmark$ ($\checkmark$ \slash\ $\checkmark$) & $\checkmark$ ($\checkmark$ \slash\ 58) \\ \hline
        \ce{Ba(FeAs)2}      & 5 & $I4/mmm$       & $\checkmark$ ($\checkmark$ \slash\ $\checkmark$) & $\checkmark$ ($\checkmark$ \slash\ 27) \\ \hline
        \ce{SiO2}           & 6 & $I\bar42d$     & $\checkmark$ ($\checkmark$ \slash\ $\checkmark$) & $\times$ ($\checkmark$ \slash\ 12) \\ \hline
        \ce{VO2}            & 6 & $P4_2/mnm$     & $\checkmark$ ($\checkmark$ \slash\ $\checkmark$) & $\checkmark$ ($\checkmark$ \slash\ 38) \\ \hline
        \ce{La2CuO4}        & 7 & $I4/mmm$       & $\times$ ($\checkmark$ \slash\ $\checkmark$) & $\checkmark$ ($\checkmark$ \slash\ 33) \\ \hline
        \ce{LiPF6}          & 8 & $R\bar3$       & $\checkmark$ ($\checkmark$ \slash\ $\checkmark$) & $\checkmark$ ($\checkmark$ \slash\ 12) \\ \hline
        \ce{Al2O3}          & 10 & $R\bar3c$     & $\checkmark$ ($\checkmark$ \slash\ $\checkmark$) & $\checkmark$ ($\checkmark$ \slash\ 91) \\ \hline
        \ce{SrTiO3}         & 10 & $I4/mcm$      & $\checkmark$ ($\checkmark$ \slash\ $\checkmark$) & $\checkmark$ ($\checkmark$ \slash\ 108) \\ \hline
        \ce{CaCO3}          & 10 & $R\bar3c$     & $\checkmark$ ($\checkmark$ \slash\ $\checkmark$) & $\checkmark$ ($\checkmark$ \slash\ 105)  \\ \hline
        \ce{TiO2}           & 12 & $C2/m$        & $\times$ ($\times$ \slash\ $\checkmark$) & $\times$ ($\checkmark$ \slash\ 43) \\ \hline
        \ce{ZrO2}           & 12 & $P2_1/c$      & $\checkmark$ ($\checkmark$ \slash\ $\checkmark$) & $\checkmark$ ($\checkmark$ \slash\ 44) \\ \hline
        \ce{ZrTe5}          & 12 & $Cmcm$        & $\checkmark$ ($\checkmark$ \slash\ $\checkmark$) & $\checkmark$ ($\checkmark$ \slash\ 32) \\ \hline
        \ce{V2O5}           & 14 & $Pmmn$        & $\checkmark$ ($\checkmark$ \slash\ $\checkmark$) & $\times$ ($\times$ \slash\ 43) \\ \hline
        \ce{Si3N4}          & 14 & $P6_3/m$      & $\checkmark$ ($\checkmark$ \slash\ $\checkmark$) & $\checkmark$ ($\checkmark$ \slash\ 43) \\ \hline
        \ce{Fe3O4}          & 14 & $Fd\bar3m$    & $\checkmark$ ($\checkmark$ \slash\ $\checkmark$) & $\checkmark$ ($\checkmark$ \slash\ 47) \\ \hline
        \ce{Mn(FeO2)2}      & 14 & $Fd\bar3m$    & $\checkmark$ ($\checkmark$ \slash\ $\checkmark$) & $\checkmark$ ($\checkmark$ \slash\ 135) \\ \hline
        \ce{ZnSb}           & 16 & $Pbca$        & $\checkmark$ ($\checkmark$ \slash\ $\checkmark$) & $\checkmark$ ($\checkmark$ \slash\ 67) \\ \hline
        \ce{CoSb3}          & 16 & $Im\bar3$     & $\checkmark$ ($\checkmark$ \slash\ $\checkmark$) & $\checkmark$ ($\checkmark$ \slash\ 4) \\ \hline
        \ce{LiBF4}          & 18 & $P3_121$      & $\checkmark$ ($\checkmark$ \slash\ $\checkmark$) & $\checkmark$ ($\checkmark$ \slash\ 26) \\ \hline
        \ce{Y2Co17}         & 19 & $R\bar3m$     & $\checkmark$ ($\checkmark$ \slash\ $\checkmark$) & $\checkmark$ ($\checkmark$ \slash\ 2) \\ \hline
        \ce{GeH4}           & 20 & $P2_12_12_1$  & $\checkmark$ ($\checkmark$ \slash\ $\checkmark$) & $\checkmark$ ($\checkmark$ \slash\ 48) \\ \hline
        \ce{CsPbI3}         & 20 & $Pnma$        & $\times$ ($\times$ \slash\ $\checkmark$)         & $\checkmark$ ($\checkmark$ \slash\ 170) \\ \hline
        \ce{NaCaAlPHO5F2}   & 24 & $P2_1/m$      & $\times$ ($\times$ \slash\ $\checkmark$)         & $-$ \\ \hline
        \ce{LiFePO4}        & 28 & $Pnma$        & $\checkmark$ ($\checkmark$ \slash\ $\checkmark$) & $\checkmark$ ($\checkmark$ \slash\ 107) \\ \hline
        \ce{Cu12Sb4S13}     & 29 & $I\bar43m$    & $\checkmark$ ($\checkmark$ \slash\ $\checkmark$) & $\checkmark$ ($\checkmark$ \slash\ 2) \\ \hline
        \ce{MgB7}           & 32 & $Imma$        & $\times$ ($\checkmark$ \slash\ $\checkmark$)     & $\times$ ($\times$ \slash\ 9) \\ \hline
        \ce{Li3PS4}         & 32 & $Pnma$        & $\checkmark$ ($\checkmark$ \slash\ $\checkmark$) & $\times$ ($\times$ \slash\ 64) \\ \hline
        \ce{Cd3As2}         & 80 & $I4_1/acd$    & $\checkmark$ ($\checkmark$ \slash\ $\checkmark$) & $\checkmark$ ($\checkmark$ \slash\ 32) \\ \hline
        \ce{Li4Ti5O12}      & 42 & $C2/c$        & $\checkmark$ ($\checkmark$ \slash\ $\checkmark$) & $\checkmark$ ($\checkmark$ \slash\ 12) \\ \hline
        \ce{Ba2CaSi4(BO7)2} & 46 & $I\bar42m$    & $\times$ ($\times$ \slash\ $\checkmark$)         & $\times$ ($\times$ \slash\ 12) \\ \hline
        \ce{Ag8GeS6}        & 60 & $Pna2_1$      & $\times$ ($\checkmark$ \slash\ $\checkmark$)     & $\checkmark$ ($\checkmark$ \slash\ 7) \\ \hline
        \ce{Nd2Fe14B}       & 68 & $P4_2/mnm$    & $\times$ ($\times$ \slash\ $\checkmark$)         & $\checkmark$ ($\checkmark$ \slash\ 2) \\ \hline
        \ce{Y3Al5O12}       & 80 & $Ia\bar3d$    & $\checkmark$ ($\checkmark$ \slash\ $\checkmark$) & $\checkmark$ ($\checkmark$ \slash\ 11) \\ \hline
        \ce{Ca14MnSb11}     & 104 & $I4_1/acd$   & $\times$ ($\checkmark$ \slash\ $\checkmark$)     & $\checkmark$ ($\checkmark$ \slash\ 2) \\ \hline
        \wcline{1-5}
        \multicolumn{1}{r}{\textbf{Overall}}        & & & $31 / 40 = 77.5\%$ & $33 / 40 = 82.5\%$ \\
    \bottomrule
\end{longtable}
% }

\begin{longtable}[!htbp]{r|ccccc}
\caption{
Results of CSP algorithm with ShotgunCSP-GT and ShotgunCSP-GW generators for the 50 crystals comprising Dataset II. See the caption of Table \ref{tab:dataset_I_results} for details.
}

\label{tab:dataset_II_results}\\
\wcline{1-5}
    \thead{Composition} & \thead{Number of atoms} & \thead{Space group} & \thead{ShotgunCSP-GW} & \thead{ShotgunCSP-GT} \\
\cline{1-5}
\endfirsthead 

\multicolumn{5}{l}{Table \ref{tab:dataset_II_results} continued}\\
\wcline{1-5}
    \thead{Composition} & \thead{Number of atoms} & \thead{Space group} & \thead{ShotgunCSP-GW} & \thead{ShotgunCSP-GT} \\
\cline{1-5}
\endhead

\wcline{1-5} 
\endfoot

\endlastfoot
    \ce{CsCl}           & 2 & $Fm\bar3m$    & $\checkmark$ ($\checkmark$ \slash\ $\checkmark$)   & $\checkmark$ ($\checkmark$ \slash\ 52) \\ \hline
    \ce{MnAl}           & 2 & $P4/mmm$      & $\checkmark$ ($\checkmark$ \slash\ $\checkmark$)   & $\checkmark$ ($\checkmark$ \slash\ 37) \\ \hline
    \ce{HoHSe}          & 3 & $P\bar6m2$    & $\checkmark$ ($\checkmark$ \slash\ $\checkmark$)   & $\checkmark$ ($\checkmark$ \slash\ 71) \\ \hline
    \ce{ErCdRh2}        & 4 & $Fm\bar3m$    & $\checkmark$ ($\checkmark$ \slash\ $\checkmark$)   & $\checkmark$ ($\checkmark$ \slash\ 113) \\ \hline
    \ce{Eu2MgTl}        & 4 & $Fm\bar3m$    & $\checkmark$ ($\checkmark$ \slash\ $\checkmark$)   & $\checkmark$ ($\checkmark$ \slash\ 114) \\ \hline
    \ce{Pm2NiIr}        & 4 & $Fm\bar3m$    & $\checkmark$ ($\checkmark$ \slash\ $\checkmark$)   & $\checkmark$ ($\checkmark$ \slash\ 113) \\ \hline
    \ce{VPt3}           & 4 & $I4/mmm$      & $\checkmark$ ($\checkmark$ \slash\ $\checkmark$)   & $\checkmark$ ($\checkmark$ \slash\ 60) \\ \hline
    \ce{Gd(SiOs)2}      & 5 & $I4/mmm$      & $\checkmark$ ($\checkmark$ \slash\ $\checkmark$)   & $\checkmark$ ($\checkmark$ \slash\ 28) \\ \hline
    \ce{LaAl3Au}        & 5 & $I4mm$        & $\checkmark$ ($\checkmark$ \slash\ $\checkmark$)   & $\checkmark$ ($\checkmark$ \slash\ 44) \\ \hline
    \ce{U2SbN2}         & 5 & $I4/mmm$      & $\checkmark$ ($\checkmark$ \slash\ $\checkmark$)   & $\checkmark$ ($\checkmark$ \slash\ 120) \\ \hline
    \ce{MnGa(CuSe2)2}   & 8 & $I\bar4$      & $\checkmark$ ($\checkmark$ \slash\ $\checkmark$)   & $\checkmark$ ($\checkmark$ \slash\ 11) \\ \hline
    \ce{SmZnPd}         & 9 & $P\bar62m$    & $\checkmark$ ($\checkmark$ \slash\ $\checkmark$)   & $\checkmark$ ($\checkmark$ \slash\ 2) \\ \hline
    \ce{Sn(TePd3)2}     & 9 & $I4mm$        & $\times$ ($\times$ \slash\ $\times$)               & $\checkmark$ ($\checkmark$ \slash\ 96) \\ \hline
    \ce{V5S4}           & 9 & $I4/m$        & $\checkmark$ ($\checkmark$ \slash\ $\checkmark$)   & $\checkmark$ ($\checkmark$ \slash\ 17) \\ \hline
    \ce{Cs3InF6}        & 10 & $Fm\bar3m$   & $\checkmark$ ($\checkmark$ \slash\ $\checkmark$)   & $\checkmark$ ($\checkmark$ \slash\ 6) \\ \hline
    \ce{Eu(CuSb)2}      & 10 & $P4/nmm$     & $\checkmark$ ($\checkmark$ \slash\ $\checkmark$)   & $\checkmark$ ($\checkmark$ \slash\ 26) \\ \hline
    \ce{Rb2TlAgCl6}     & 10 & $Fm\bar3m$   & $\checkmark$ ($\checkmark$ \slash\ $\checkmark$)   & $\checkmark$ ($\checkmark$ \slash\ 3) \\ \hline
    \ce{Ca3Ni7B2}       & 12 & $R\bar3m$    & $\checkmark$ ($\checkmark$ \slash\ $\checkmark$)   & $\checkmark$ ($\checkmark$ \slash\ 18) \\ \hline
    \ce{DyPO4}          & 12 & $I4_1/amd$   & $\checkmark$ ($\checkmark$ \slash\ $\checkmark$)   & $\checkmark$ ($\checkmark$ \slash\ 138) \\ \hline
    \ce{LaSiIr}         & 12 & $P2_13$      & $\checkmark$ ($\checkmark$ \slash\ $\checkmark$)   & $\checkmark$ ($\checkmark$ \slash\ 33) \\ \hline
    \ce{SmVO4}          & 12 & $I4_1/amd$   & $\checkmark$ ($\checkmark$ \slash\ $\checkmark$)   & $\checkmark$ ($\checkmark$ \slash\ 136) \\ \hline
    \ce{VCl5}           & 12 & $P\bar1$     & $\checkmark$ ($\checkmark$ \slash\ $\checkmark$)       & $\checkmark$ ($\checkmark$ \slash\ 32) \\ \hline
    \ce{YbP5}           & 12 & $P2_1/m$     & $\times$ ($\checkmark$ \slash\ $\checkmark$)       & $\checkmark$ ($\checkmark$ \slash\ 3) \\ \hline
    \ce{Eu(Al2Cu)4}     & 13 & $I4/mmm$     & $\checkmark$ ($\checkmark$ \slash\ $\checkmark$)   & $\checkmark$ ($\checkmark$ \slash\ 1) \\ \hline
    \ce{Zr4O}           & 15 & $R\bar3$     & $\times$ ($\times$ \slash\ $\times$)               & $\times$ ($\times$ \slash\ 16)         \\ \hline
    \ce{Ba3Ta2NiO9}     & 15 & $P\bar3m1$   & $\checkmark$ ($\checkmark$ \slash\ $\checkmark$)   & $\times$ ($\times$ \slash\ 39)         \\ \hline
    \ce{K2Ni3S4}        & 18 & $Fddd$       & $\checkmark$ ($\checkmark$ \slash\ $\checkmark$)   & $\checkmark$ ($\checkmark$ \slash\ 78) \\ \hline
    \ce{Sr(ClO3)2}      & 18 & $Fdd2$       & $\checkmark$ ($\checkmark$ \slash\ $\checkmark$)   & $\checkmark$ ($\checkmark$ \slash\ 76) \\ \hline
    \ce{LiSm2IrO6}      & 20 & $P2_1/c$     & $\times$ ($\checkmark$ \slash\ $\checkmark$)       & $\checkmark$ ($\checkmark$ \slash\ 83) \\ \hline
    \ce{Pr2ZnPtO6}      & 20 & $P2_1/c$     & $\times$ ($\checkmark$ \slash\ $\checkmark$)       & $\checkmark$ ($\checkmark$ \slash\ 85) \\ \hline
    \ce{Sc2Mn12P7}      & 21 & $P\bar6$     & $\checkmark$ ($\checkmark$ \slash\ $\checkmark$)   & $\checkmark$ ($\checkmark$ \slash\ 7) \\ \hline
    \ce{LaSi2Ni9}       & 24 & $I4_1$/amd   & $\checkmark$ ($\checkmark$ \slash\ $\checkmark$)   & $\checkmark$ ($\checkmark$ \slash\ 2) \\ \hline
    \ce{CeCu5Sn}        & 28 & $Pnma$       & $\checkmark$ ($\checkmark$ \slash\ $\checkmark$)   & $\checkmark$ ($\checkmark$ \slash\ 2) \\ \hline
    \ce{LiP(HO2)2}      & 32 & $Pna2_1$     & $\times$ ($\checkmark$ \slash\ $\checkmark$)       & $\checkmark$ ($\times$ \slash\ 99)     \\ \hline
    \ce{Mg3Si2H4O9}     & 36 & $P6_3cm$     & $\checkmark$ ($\times$ \slash\ $\times$)               & $\times$ ($\times$ \slash\ 1)         \\ \hline
    \ce{Y4Si5Ir9}       & 36 & $P6_3/mmc$   & $\checkmark$ ($\checkmark$ \slash\ $\checkmark$)   & $-$                         \\ \hline
    \ce{Na(WO3)9}       & 37 & $R\bar3$     & $\checkmark$ ($\checkmark$ \slash\ $\checkmark$)   & $-$                         \\ \hline
    \ce{Sm6Ni20As13}    & 39 & $P\bar6$     & $\checkmark$ ($\checkmark$ \slash\ $\checkmark$)   & $\checkmark$ ($\checkmark$ \slash\ 2) \\ \hline
    \ce{BaCaGaF7}       & 40 & $P2/c$       & $\checkmark$ ($\checkmark$ \slash\ $\checkmark$)       & $\checkmark$ ($\checkmark$ \slash\ 6) \\ \hline
    \ce{Tm11Sn10}       & 42 & $I4/mmm$     & $\checkmark$ ($\checkmark$ \slash\ $\checkmark$)   & $\checkmark$ ($\checkmark$ \slash\ 3) \\ \hline
    \ce{AlH12(ClO2)3}   & 44 & $R\bar3c$    & $\checkmark$ ($\times$ \slash\ $\times$)               & $\checkmark$ ($\checkmark$ \slash\ 11) \\ \hline
    \ce{K2ZrSi2O7}      & 48 & $P2_1/c$     & $\checkmark$ ($\checkmark$ \slash\ $\checkmark$)       & $\checkmark$ ($\times$ \slash\ 32)     \\ \hline
    \ce{LiZr2(PO4)3}    & 72 & $P2_1/c$     & $\times$ ($\times$ \slash\ $\checkmark$)           & $\checkmark$ ($\checkmark$ \slash\ 21) \\ \hline
    \ce{K5Ag2(AsSe3)3}  & 76 & $Pnma$       & $\checkmark$ ($\checkmark$ \slash\ $\checkmark$)   & $-$                         \\ \hline
    \ce{Be17Ru3}        & 80 & $Im\bar3$    & $\checkmark$ ($\times$ \slash\ $\checkmark$)           & $\checkmark$ ($\checkmark$ \slash\ 1) \\ \hline
    \ce{Cu3P8(S2Cl)3}   & 80 & $Pnma$       & $\times$ ($\times$ \slash\ $\checkmark$)           & $\checkmark$ ($\checkmark$ \slash\ 2) \\ \hline
    \ce{Al2CoO4}        & 84 & $P3m1$       & $\times$ ($\times$ \slash\ $\times$)               & $\checkmark$ ($\checkmark$ \slash\ 15) \\ \hline
    \ce{Li6V3P8O29}     & 92 & $P1$         & $\times$ ($\checkmark$ \slash\ $\checkmark$)   & $\checkmark$ ($\times$ \slash\ 5)     \\ \hline
    \ce{ReBi3O8}        & 96 & $P2_13$      & $\checkmark$ ($\checkmark$ \slash\ $\checkmark$)   & $\checkmark$ ($\checkmark$ \slash\ 7) \\ \hline
    \ce{Na5FeP2(O4F)2}  & 288 & $Pbca$      & $\times$ ($\times$ \slash\ $\checkmark$)           & $-$                         \\ \hline
\wcline{1-5}
\multicolumn{1}{r}{\textbf{Overall}}        & & & $39 / 50 = 78.0\%$ & $43 / 50 = 86.0\%$ \\

\bottomrule
\end{longtable}

\begin{figure}[!htbp]
\centering
\includegraphics[width=0.95\linewidth]{structure_visualization.pdf}
\caption{Examples of crystal structures predicted by the proposed CSP algorithms (depicted with VESTA \cite{momma2011VESTAThreedimensional} version 3.5.8). For each generation method (ShotgunCSP-GT and ShotgunCSP-GW), the predicted structures with the two lowest DFT energies are shown. The true (target) stable structures are shown on the left.}
\label{fig:4}
\end{figure}

\section*{Discussion}
This paper presents a CSP workflow based on a machine-learning approach for the efficient prediction of stable crystal structures without iterative DFT calculations. The essence of the proposed method is a shotgun-type virtual screening of crystal structures. A surrogate model that predicts DFT energies is used to screen a large number of virtual crystal structures. The efficiently narrowed-down candidate structures are then relaxed using DFT calculations to predict stable crystal structures. The key technical components in this workflow are the surrogate model for energy prediction and the crystal structure generators. To train the surrogate model for DFT energy calculations, a pretrained CGCNN is fine-tuned for the prediction of the energies of virtual crystal structures in pre-relaxed states to decrease the number of training samples generated with DFT single-point energy calculations. Virtual libraries of candidate crystal structures are constructed by either element substitution of template crystals  (ShotgunCSP-GT) or a Wyckoff generator involving space-group prediction and Wyckoff-label prediction  (ShotgunCSP-GW). Of 90 known crystal structures (Datasets I and II) with a wide range of chemical compositions, symmetries, and structure types, the ShotgunCSP-GT- and ShotgunCSP-GW-based workflows successfully predicted 84.4\% and 74.4\% of the true structures, respectively.

For the 25 benchmark structures successfully predicted by USPEX, our ShotgunCSP-GW-based method achieved the same or better prediction accuracy and reduced the computation time by approximately 40\%. To our knowledge, our method is the simplest CSP algorithm available today. A significant contribution of this study is in proving that such a straightforward approach can effectively predict numerous crystal structures that cannot be predicted using conventional methods, including those with low symmetry and large-unit-cell systems. However, our method remains incapable of predicting more highly complex crystal systems, such as those in Dataset III, which were selected to be more challenging. The bottleneck is expected to be partially eliminated by achieving the efficient prediction of the true Wyckoff labels of large and complicated crystalline systems.

Nonetheless, the simplicity of the proposed makes it well-suited for parallel computing and managing a much larger database of candidate structures.  Moreover, it can be integrated with deep generative models for crystal structure prediction, which are expected to advance further in the future.

\section*{Methods}

\subsection*{ShotgunCSP-GT: template-based structure generation by element substitution}

The calculation procedure for template-based structure generation by element substitution is as follows:
\begin{enumerate}
\setlength{\itemsep}{0cm}
\item Extract template structures with the same composition ratio as the query composition $X$ from 33,040 stable structures in the Materials Project database.
\item Replace elements in the templates with elements that have the same number of atoms in $X$. If the substitution target is not uniquely determined, substitute the element that has the smallest Euclidean distance in XenonPy's 58-dimensional element descriptors.
\item Convert the chemical compositions of the template structures to the 290-dimensional descriptors in XenonPy and apply DBSCAN clustering to group the template structures.
\item Extract template sets that belong to the same group as $X$. Furthermore, using the StructureMatcher module of pymatgen \cite{ong2013PythonMaterials, StructureMatcher}, remove structurally redundant templates to obtain a unique template set (where $K_{\mathrm{temp}}$ is the number of unique templates in the same cluster as the query composition).
\item Estimate the lattice constants using a model that predicts the unit-cell volume from the composition $X$ (see Supplementary Information).
\item Add perturbations to the atomic coordinates of each template with additive noise following the uniform distribution $U(-0.05, 0.05)$.
\item Add perturbations to the unit-cell volumes of each template with additive noise following the uniform distribution $U(-0.1, 0.1)$.
\end{enumerate}

\subsection*{ShotgunCSP-GW: Wyckoff position generator}

The calculation procedure for the Wyckoff position generator is as follows:
\begin{enumerate}
\setlength{\itemsep}{0cm}
\item Predict the space group and probabilities of Wyckoff letters of the query composition $X=X_{c_1}^1 X_{c_2}^2 \cdots X_{c_K}^K$ (where $X^k$ denotes a chemical element $k = 1, \ldots, K$ and $c^k$ is its composition ratio) for each predicted space group.
\item Extract the set ${W} = \{(l_i, m_i) | i=1,\ldots,j \}$ of Wyckoff letters $l_i$ and multiplicity $m_i$ for a predicted space group.
\item Randomly sample an element $X^k$ from composition $X$ and a possible Wyckoff letter $l_i$ with the predicted Wyckoff-letter-configuration probability $p_i$ from set ${W}$, then assign $l_i$ to $X^k$ with its possible multiplicity $m_i$.
\item Remove the assigned atoms from composition $X$ and define the remaining composition as new $X$.
\item Remove the used Wyckoff letter $l_i$ from the set ${W}$ if the Wyckoff position is exclusive, and re-normalize the probability of the remaining Wyckoff letters.
\begin{eqnarray*}
p_i \leftarrow p_i / \sum{p_i} 
\end{eqnarray*}
\item Repeat steps 3--5 until all atoms are assigned.
\item Determine the fractional coordinates of atomic sites to which the same Wyckoff letter is assigned. If the coordinates of the Wyckoff position $(x, y, z)$ of the atomic sites are allowed to vary, sample each coordinate position from a uniform distribution $U(0, 1)$.
\item Estimate the lattice constants using a model that predicts the volume from the composition $X$ (see Supplementary Information). Add perturbations to the unit-cell volume of each template with additive noise following the uniform distribution $U(-0.1, 0.1)$.
\end{enumerate}

\subsection*{Compositional descriptor}
The chemical formula of the query composition is $X = X^1_{c^1}X^2_{c^2} \cdots X^K_{c^K}$, where $X^k$ denotes a chemical element $k = 1, \ldots, K$ and $c^k$ is its composition ratio. Each element of the descriptor vector of length 290 takes the following form:
\begin{eqnarray}
\phi_{g, \eta}(X) = g(c^1, \ldots, c^K, \eta(X^1), \ldots, \eta(X^K)). 
\end{eqnarray}
The scalar quantity $\eta(X^k)$ on the right-hand side of Eq. (2) (where $k = 1, \ldots, K$) represents a feature value of the element $X^k$, such as the atomic weight, electronegativity, or polarizability. Using function $g$, the element features $\eta(X^1), \ldots, \eta(X^K)$ with compositions $c^1, \ldots, c^K$ are converted into a compositional feature. For $g$, we used five different summary statistics: weighted mean $\phi_{\mathrm{ave}}$, weighted variance $\phi_{\mathrm{var}}$, weighted sum $\phi_{\mathrm{sum}}$, max-pooling $\phi_{\mathrm{max}}$, and min-pooling $\phi_{\mathrm{min}}$:
\begin{eqnarray}
& & \phi_{\mathrm{ave}, \eta}(X) = \frac{1}{\sum_{k=1}^K c^k } \sum_{k=1}^K c^k \eta(X^k), \nonumber \\
& & \phi_{\mathrm{var, \eta}}(X) = \frac{1}{\sum_{k=1}^K c^k} \sum_{k=1}^K c^k (\eta(X^k) - \phi_{\mathrm{ave}, \eta}(X))^2, \nonumber \\
& & \phi_{\mathrm{sum, \eta}}(X) = \sum_{k=1}^K c^k \eta(X^k), \nonumber \\
&&\phi_{\max, \eta}(X) = \max\{\eta(X^1), \ldots, \eta(X^K)\}, \nonumber \\
&&\phi_{\min, \eta}(X) = \min\{\eta(X^1), \ldots, \eta(X^K)\}. \nonumber 
\end{eqnarray}
We used 58 distinct elemental features implemented in XenonPy, including the atomic number, covalent radius, van der Waals radius, electronegativity, thermal conductivity, band gap, polarizability, boiling point, and melting point. The full list of 58 features is summarized by Liu et al. \cite{liu2021MachineLearning}. In summary, composition $X$ is characterized by a 290-dimensional descriptor vector ($=58 \times 5$).

\subsection*{Fine-tuning of CGCNN}
A CGCNN localized to the energy prediction of a specific system with composition $X$ was obtained by fine-tuning the pretrained global CGCNN model from Xie and Grossman \cite{xie2018CrystalGraph} with randomly generated crystalline conformations and their DFT formation energies. We generated 100 training crystal structures for each candidate space group using the ShotgunCSP-GW generator or ten random structures from each selected template using the ShotgunCSP-GT generator. The pretrained model without the output layer was copied to the target model, to which a new output layer was subsequently added, and its parameters were randomly initialized. We then trained the target model on the target dataset. The hyperparameters, including the learning rate and gradient clipping value, were optimized by performing a grid search with the same range $\{0.01, 0.008, 0.006, 0.004, 0.002\}$, with early stopping based on the MAE of the validation set. The maximum number of epochs was fixed at 350.

\subsection*{DFT calculations}

All DFT calculations were performed using VASP (version 6.1.2) \cite{kresse1996EfficientIterative} with PAW pseudopotentials \cite{blochl1994ProjectorAugmentedwave}. The Perdew--Burke--Ernzerhof exchange--correlation functional \cite{perdew1996GeneralizedGradient} was considered for generalized gradient approximation. Brillouin zone integration of the unit cells was automatically determined using the $\Gamma$-centered Monkhorst--Pack mesh function implemented in the VASP code. Single-point energy calculations (also known as self-consistent field calculations) were performed on unrelaxed crystal structures that were created virtually to produce a training set for fine-tuning the pretrained CGCNN. The geometry of the final selected candidate structure was locally optimized by performing DFT calculations. We used the MPStaticSet and MPRelaxSet presets implemented in pymatgen \cite{ong2013PythonMaterials} with significant modifications to generate the inputs for all VASP calculations (see Supplementary Information).

\subsection*{Structural similarity}

To calculate the similarity between two structures, we encoded each query into a vector-type structural descriptor with its local coordination information (site fingerprint) from all sites \cite{zimmermann2017AssessingLocal}. Then, the structural similarity $\tau$ was calculated as the Euclidean distance between the two vectorized crystal structures. Note that the descriptor does not contain any information about the element species. The calculations were performed in matminer \cite{ward2018MatminerOpen}, which is an open-source toolkit for materials data mining, with the same configuration as that used officially in the Materials Project database. We also visually inspected the differences between structures with different $\tau$. Structures with a dissimilarity of $\tau \leq 0.2$ were treated as similar structures.

\subsection*{USPEX calculations}

USPEX calculations were performed using the official USPEX package (version 10.5) \cite{USPEXDownlaod}. We specified the calculation parameters \texttt{calculationMethod}, \texttt{calculationType}, and \texttt{optType} as ``USPEX,'' ``300,'' and ``enthalpy'', respectively, to perform the CSP task for bulk crystals using an evolutionary algorithm. The related parameters were specified according to official recommendations \cite{USPEXManual}. For example, the number of structures in each generation was set to $2 \times N$ rounded to the nearest ten, where $N$ is the number of atoms. The calculation was terminated as soon as the best structure remained unchanged over $M$ generations, where $M = N$ (rounded to the nearest ten; see Supplementary Information). Structure relaxation was executed automatically in VASP (version 6.1.2) combined with the PAW pseudopotentials. The VASP calculation settings were the same as those described in the DFT calculations section.

\section*{Data availability}
The benchmark datasets are published on Figshare and can be accessed via the following DOI: \url{https://doi.org/10.6084/m9.figshare.26536375}. Additional data supporting the findings of this study are available from the corresponding author upon reasonable request.

\section*{Code availability}
The generator of ShotgunCSP-GT is built as a Python package published on the GitHub website \url{https://github.com/TsumiNa/ShotgunCSP}.

\section*{Acknowledgements}
This work was supported in part by a Ministry of Education, Culture, Sports, Science and Technology (MEXT) KAKENHI Grant-in-Aid for Scientific Research on Innovative Areas (grant number 19H05820); Japan Society for the Promotion of Science (JSPS) Grants-in-Aid for Scientific Research (A) (grant number 19H01132) and Early-Career Scientists (grant number 23K16955); and JST CREST (grant numbers JPMJCR19I3, JPMJCR22O3, and JPMJCR2332). Computational resources were partly provided by the supercomputer at the Research Center for Computational Science, Okazaki, Japan (projects 23-IMS-C113 and 24-IMS-C107).

\section*{Author Contributions}
R.Y. and H.T. designed and conceived the project and R.Y. wrote the preliminary draft of the paper. R.Y. and C.L. designed and developed the machine-learning framework. C.L. developed the software and performed the experiments with the support of H.T., T.Y., K.W., S.Y., and A.R.O.. H.T. and S.Y. designed and tested the benchmark crystal structures. R.Y., C.L., H.T., and T.Y. wrote and revised the manuscript. All authors discussed the results and commented on the manuscript.

\section*{Additional Information}
Conflicts of Interest: The authors declare no competing interests.

\clearpage
% \FloatBarrier

%% references

\end{document}

% \linenumbers

\flushbottom
\maketitle

\clearpage

\tableofcontents

\listoftables

\listoffigures

\clearpage

\section*{Benchmark datasets}
\addcontentsline{toc}{section}{Benchmark datasets}

\begin{figure}[!ht]
\centering
\includegraphics[width=0.95\linewidth]{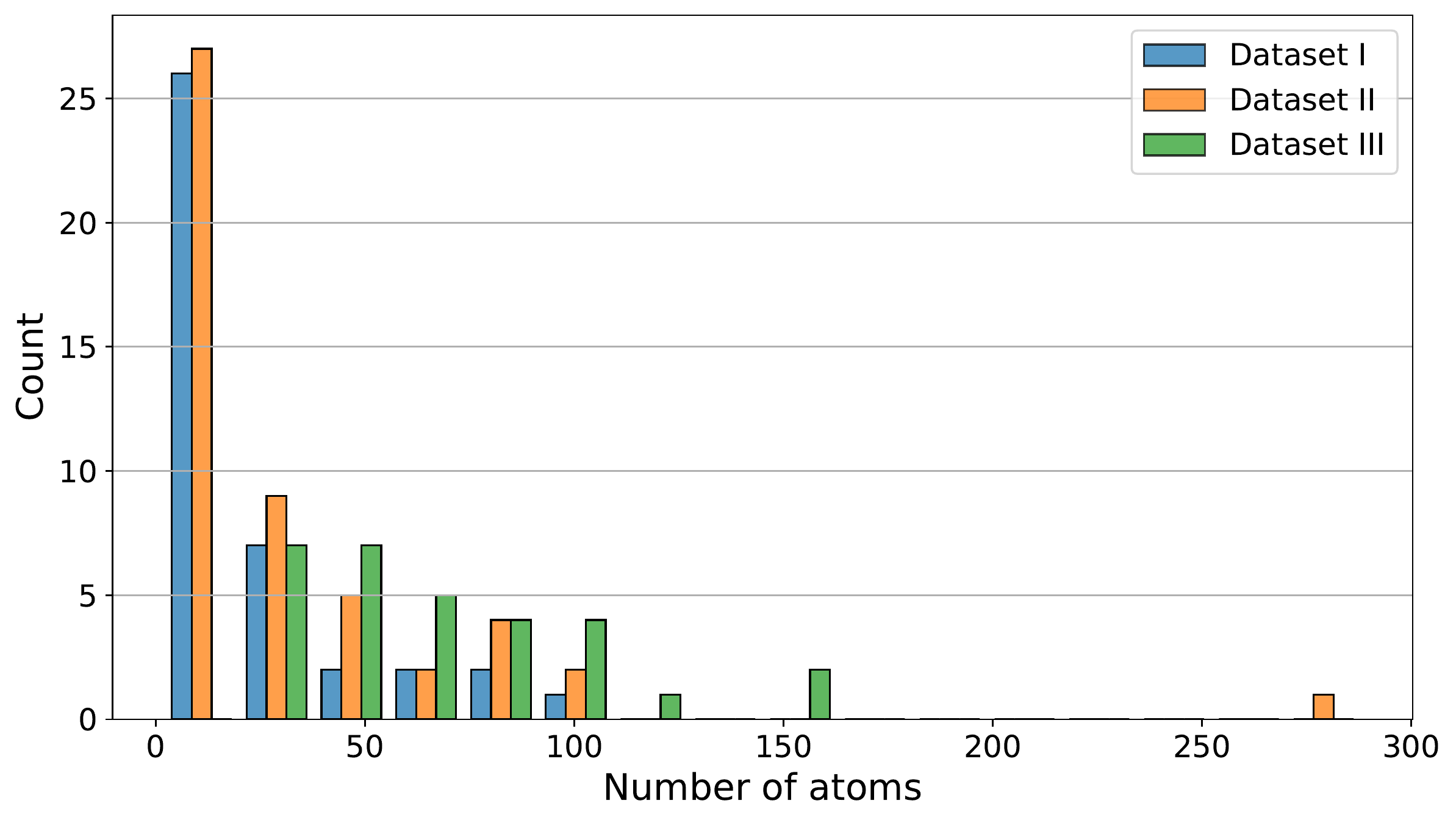}
\caption{Numbers of atoms in the unit cells of crystal structures in Datasets I, II, and III.}
\label{fig:S1_hist}
% \addtocounter{figure}{1}
\end{figure}

Dataset I consists of 40 stable crystal structures selected based on criteria including the diversity of the number of atoms in the unit cells and space groups. Dataset II consists of 50 stable crystal structures randomly selected from the Materials Project database. The chemical compositions and space groups are listed in Tables 1 and 2 in the main text, and the crystal structures are visualized in Figure \ref{fig:S4_structure}. Figure \ref{fig:S1_hist} shows histograms of the numbers of atoms in the crystal structures of Datasets I and II. The numbers of atoms ranged from 2 to 104 and from 2 to 288, respectively, with means $\pm$ standard deviations of $23.13 \pm 24.09$ and $32.68 \pm 45.41$, respectively. Dataset III consists of 30 randomly selected stable structures for which no template exists in the Materials Project database. As is evident from Table \ref{table:S1_dataset_III}, most of the crystal structures in Dataset III have far more atoms in their unit cells than those in Datasets I and II, ranging from 22 to 152 (average of $66.50 \pm 34.40$).

\renewcommand\theadalign{cc}
\renewcommand\theadfont{\bfseries}
\renewcommand\theadgape{\Gape[3pt]}
\renewcommand\cellgape{\Gape[4pt]}

\begin{table}[!ht]
\centering
\caption{Performance of the CSP algorithm with ShotgunCSP-GW for 30 randomly selected stable crystals comprising Dataset III. No template exists for these crystal structures in the Materials Project database. The top ten virtual structures with the lowest DFT energies were proposed as the final candidate. In the fourth column, symbols $\checkmark$ and $\times$ indicate prediction success and failure, respectively. The left and right sides of the parentheses indicate the success or failure of Wyckoff pattern prediction and space-group prediction, respectively.
}
\begin{tabular}{r|ccc}
\wcline{1-4}
\thead{Composition} & \thead{Number of atoms} & \thead{Space group} & \thead{CSP using\\Wyckoff position\\generator} \\
\cline{1-4}
    \ce{Ba3Si6N4O9} & 22 & $P3$ & $\times$ ($\times$ \slash\ $\times$) \\ \hline
    \ce{Ho10Te7S10} & 27 & $P1$ & $\times$ ($\times$ \slash\ $\checkmark$) \\ \hline
    \ce{Ti13Al9Co8} & 30 & $R3m$ & $\times$ ($\times$ \slash\ $\checkmark$) \\ \hline
    \ce{K11LiMn4O16} & 32 & $I\bar42m$ & $\times$ ($\times$ \slash\ $\checkmark$) \\ \hline
    \ce{BaSrMn2Al9PbO20} & 34 & $P1$ & $\times$ ($\times$ \slash\ $\checkmark$) \\ \hline
    \ce{RbNa3Li12Ti4O16} & 36 & $I4/m$ & $\times$ ($\times$ \slash\ $\times$) \\ \hline
    \ce{B10(Pb2O7)3} & 37 & $P\bar1$ & $\times$ ($\times$ \slash\ $\checkmark$) \\ \hline
    \ce{Nb12Br17F13} & 42 & $P1$ & $\times$ ($\times$ \slash\ $\checkmark$) \\ \hline
    \ce{Na4PuH7O9} & 42 & $P\bar1$ & $\times$ ($\times$ \slash\ $\checkmark$) \\ \hline
    \ce{RbLa2C6N6ClO6} & 44 & $P6_3/m$ & $\times$ ($\times$ \slash\ $\checkmark$) \\ \hline
    \ce{Mg10B16Ir19} & 45 & $I\bar43m$ & $\checkmark$ ($\checkmark$ \slash\ $\checkmark$) \\ \hline
    \ce{LiMn3Al2(HO2)6} & 48 & $P\bar1$ & $\times$ ($\times$ \slash\ $\checkmark$) \\ \hline
    \ce{Na8Al6Si6CO27} & 48 & $R3$ & $\times$ ($\times$ \slash\ $\times$) \\ \hline
    \ce{Sr16V8O31} & 55 & $P\bar1$ & $\times$ ($\times$ \slash\ $\checkmark$) \\ \hline
    \ce{Nd14Zn43Sn3} & 60 & $Pm$ & $\times$ ($\times$ \slash\ $\checkmark$) \\ \hline
    \ce{H18SO12} & 62 & $Cc$ & $\times$ ($\times$ \slash\ $\checkmark$) \\ \hline
    \ce{Cs10(Mo2N5)3} & 62 & $R\bar3c$ & $\times$ ($\times$ \slash\ $\checkmark$) \\ \hline
    \ce{La17Al4(Si3N11)3} & 63 & $F\bar43m$ & $\times$ ($\times$ \slash\ $\times$) \\ \hline
    \ce{Ba2V5(PO6)4} & 70 & $Cm$ & $\times$ ($\times$ \slash\ $\times$) \\ \hline
    \ce{Ba10Li2Bi4O21} & 74 & $Pmc2_1$ & $\times$ ($\times$ \slash\ $\checkmark$) \\ \hline
    \ce{Bi8AsAuCl9} & 76 & $P2_1/c$ & $\times$ ($\times$ \slash\ $\checkmark$) \\ \hline
    \ce{Ba8Nb7S24} & 78 & $P2_1/c$ & $\times$ ($\times$ \slash\ $\checkmark$) \\ \hline
    \ce{KBa6Zn4(GaO3)7} & 78 & $P1$ & $\times$ ($\times$ \slash\ $\times$) \\ \hline
    \ce{Al2Si3H8(NO5)2} & 100 & $P2_1$ & $\times$ ($\times$ \slash\ $\checkmark$) \\ \hline
    \ce{H13S2N3O8} & 104 & $P2/c$ & $\times$ ($\times$ \slash\ $\times$) \\ \hline
    \ce{KNa3Al12H24(SO7)8} & 104 & $P2/m$ & $\times$ ($\times$ \slash\ $\times$) \\ \hline
    \ce{Er6Al41Cr6} & 106 & $P\bar31m$ & $\times$ ($\times$ \slash\ $\checkmark$) \\ \hline
    \ce{K3Fe3P4H2O17} & 116 & $Pnna$ & $\times$ ($\times$ \slash\ $\checkmark$) \\ \hline
    \ce{Ta4P4S29} & 148 & $P4_12_12$ & $\times$ ($\times$ \slash\ $\times$) \\ \hline
    \ce{Al8(Pb3O7)3} & 152 & $Pa\bar3$ & $\checkmark$ ($\checkmark$ \slash\ $\checkmark$) \\ \hline
    \wcline{1-4}
    \multicolumn{1}{r}{\textbf{Overall}}        & & & $2 / 30 = 6.7\%$ \\
\bottomrule
\end{tabular}

\label{table:S1_dataset_III}
\end{table}

\section*{Energy predictions}
\addcontentsline{toc}{section}{Energy predictions}

We employed transfer learning to fine-tune the crystal graph convolutional neural network (CGCNN), introduced by Xie et al. \cite{xie2018CrystalGraph}, for the prediction of formation energies of virtually generated crystal structures. The CGCNN was first pretrained on a substantial dataset comprising 126,210 crystal structures extracted from the Materials Project database. The 120 benchmark crystal structures were excluded from the training data. For a given composition, single-point energy calculations were performed on 1000 generated virtual structures to evaluate their energies. The energies were then used to fine-tune the localized model by transfer learning.

The energy prediction calculations were executed on the supercomputer AI Bridging Cloud Infrastructure (ABCI) \cite{abci} by utilizing 20 cores of an Intel Xeon Gold 6148 CPU per job. Table \ref{tab:S2_runing_time} provides a comprehensive summary of the computational cost associated with single-point energy calculations for the 40 cases in Dataset I, including the average, standard deviation, minimum, 25th percentile, 50th percentile, 75th percentile, and maximum computation time (measured in seconds). In most cases (e.g., for compositions with fewer than 30 atoms per unit cell), calculations were completed in less than 2 min. Calculations for compositions with larger unit cells (e.g., \ce{Y3Al5O12} and \ce{Ca14MnSb11}) took more than 30 min to run to completion.

\begin{table}[ht!]
\centering
\caption{Computation time (measured in seconds) for single-point energy calculations of Dataset I.}
\scalebox{0.95}{
\begin{tabular}{l|cccccccc}
\wcline{1-9}
\thead{Composition}    & \thead{Number\\of atoms} & \thead{Average} & \thead{Standard\\deviation} & \thead{Minimum} & \thead{25\%} & \thead{50\%} & \thead{75\%} & \thead{Maximum} \\
\cline{1-9}
\ce{C}              & 4               & 14      & 5                  & 5       & 11   & 13   & 15   & 39      \\ \hline
\ce{Si}             & 2               & 5       & 1                  & 4       & 4    & 5    & 6    & 8       \\ \hline
\ce{GaAs}           & 2               & 10      & 1                  & 7       & 9    & 10   & 10   & 15      \\ \hline
\ce{ZnO}            & 4               & 12      & 6                  & 7       & 9    & 11   & 13   & 39      \\ \hline
\ce{BN}             & 4               & 10      & 3                  & 6       & 8    & 9    & 11   & 26      \\ \hline
\ce{LiCoO2}         & 16              & 17      & 7                  & 9       & 12   & 14   & 19   & 46      \\ \hline
\ce{Bi2Te3}         & 5               & 16      & 4                  & 11      & 14   & 15   & 17   & 37      \\ \hline
\ce{Ba(FeAs)2}      & 5               & 18      & 5                  & 12      & 16   & 17   & 20   & 40      \\ \hline
\ce{SiO2}           & 6               & 207     & 133                & 90      & 118  & 162  & 243  & 749     \\ \hline
\ce{VO2}            & 6               & 33      & 16                 & 12      & 22   & 30   & 42   & 103     \\ \hline
\ce{La2CuO4}        & 7               & 51      & 27                 & 26      & 36   & 41   & 48   & 154     \\ \hline
\ce{LiPF6}          & 8               & 17      & 6                  & 9       & 12   & 15   & 20   & 39      \\ \hline
\ce{Al2O3}          & 10              & 12      & 4                  & 6       & 10   & 12   & 14   & 27      \\ \hline
\ce{SrTiO3}         & 10              & 33      & 14                 & 11      & 24   & 31   & 41   & 76      \\ \hline
\ce{CaCO3}          & 10              & 26      & 12                 & 12      & 20   & 23   & 28   & 80      \\ \hline
\ce{TiO2}           & 12              & 29      & 15                 & 11      & 20   & 27   & 34   & 128     \\ \hline
\ce{ZrO2}           & 12              & 32      & 11                 & 9       & 25   & 31   & 36   & 81      \\ \hline
\ce{ZrTe5}          & 12              & 48      & 24                 & 15      & 37   & 41   & 51   & 140     \\ \hline
\ce{V2O5}           & 14              & 45      & 18                 & 24      & 33   & 40   & 52   & 133     \\ \hline
\ce{Si3N4}          & 14              & 20      & 6                  & 10      & 17   & 19   & 22   & 41      \\ \hline
\ce{Fe3O4}          & 14              & 75      & 35                 & 14      & 50   & 69   & 90   & 173     \\ \hline
\ce{Mn(FeO2)2}      & 14              & 59      & 29                 & 28      & 40   & 50   & 64   & 180     \\ \hline
\ce{ZnSb}           & 16              & 45      & 14                 & 7       & 39   & 45   & 51   & 86      \\ \hline
\ce{CoSb3}          & 16              & 42      & 27                 & 11      & 31   & 38   & 49   & 185     \\ \hline
\ce{LiBF4}          & 18              & 66      & 31                 & 25      & 39   & 57   & 81   & 102     \\ \hline
\ce{Y2CO17}         & 19              & 187     & 22                 & 177     & 181  & 183  & 206  & 212     \\ \hline
\ce{GeH4}           & 20              & 33      & 10                 & 20      & 27   & 31   & 37   & 87      \\ \hline
\ce{CsPbI3}         & 20              & 130     & 73                 & 37      & 99   & 124  & 157  & 339     \\ \hline
\ce{NaCaAlPHO5F2}   & 24              & 124     & 33                 & 92      & 101  & 123  & 145  & 198     \\ \hline
\ce{LiFePO4}        & 28              & 164     & 89                 & 66      & 105  & 130  & 197  & 615     \\ \hline
\ce{Cu12Sb4S13}     & 29              & 134     & 26                 & 107     & 126  & 128  & 131  & 201     \\ \hline
\ce{MgB7}           & 32              & 38      & 12                 & 9       & 35   & 39   & 45   & 59      \\ \hline
\ce{Li3PS4}         & 32              & 151     & 74                 & 75      & 100  & 126  & 169  & 405     \\ \hline
\ce{Cd3As2}         & 80              & 216     & 31                 & 191     & 199  & 208  & 217  & 288     \\ \hline
\ce{Li4Ti5O12}      & 42              & 375     & 206                & 153     & 228  & 318  & 434  & 1077    \\ \hline
\ce{Ba2CaSi4(BO7)2} & 46              & 249     & 69                 & 131     & 210  & 245  & 272  & 439     \\ \hline
\ce{Ag8GeS6}        & 60              & 489     & 36                 & 453     & 461  & 466  & 517  & 544     \\ \hline
\ce{Nd2Fe14B}       & 68              & 1026    & 178                & 678     & 823  & 911  & 1223 & 1785    \\ \hline
\ce{Y3Al5O12}       & 80              & 1489    & 233                & 998     & 1221 & 1366 & 1467 & 2648    \\ \hline
\ce{Ca14MnSb11}     & 104             & 1912    & 591                & 1005    & 1451 & 1859 & 2103 & 3424    \\ \bottomrule
\end{tabular}
}
\label{tab:S2_runing_time}
\end{table}

\section*{Space-group predictor}
\addcontentsline{toc}{section}{Space-group predictor}

Using 33,040 stable crystal structures obtained from the Materials Project database and their space groups, excluding the 120 benchmark crystal structures, we constructed a fully connected neural network that classifies any composition into one of 213 space groups. The structures were split into training and test sets in a 4:1 ratio. Fivefold cross-validation was performed on the training set to identify the hyperparameters with the highest average prediction accuracy. Specifically, the hyperparameters were selected by Bayesian optimization using the Optuna Python library (the number of trials was set to 200) \cite{akiba2019OptunaNextgeneration} from the candidate set: the number of layers $\in \{2,3,4\}$, dropout ratio $\in \{0,0.1,0.2\}$, and number of neurons for each layer ($N \times \kappa$, where $N$ denotes the number of neurons in the previous layer and $\kappa \in [0.8, 0.95]$). Finally, we used the test set to calculate the predictive performance of the neural network that was trained using the hyperparameters selected by Optuna. Random data partitioning was repeated independently 100 times to determine the mean and variance of the prediction accuracy. The precision, recall, and F$_1$ scores for the top ten predicted structures were $0.8535 \pm 0.0053$, $0.8535 \pm 0.0054$, and $0.8535 \pm 0.0054$, respectively. Figure~\ref{fig:S2_recall_each_spg} shows the recall rate versus the number of training samples for each space group. The 25th, 50th, and 75th percentiles of the recall rates by space group were 53.05\%, 72.62\%, and 84.25\%, respectively. The variability of the recall rates was partially correlated with the number of training instances for each space group.

\begin{figure}[!ht]
\centering
\includegraphics[width=0.95\linewidth]{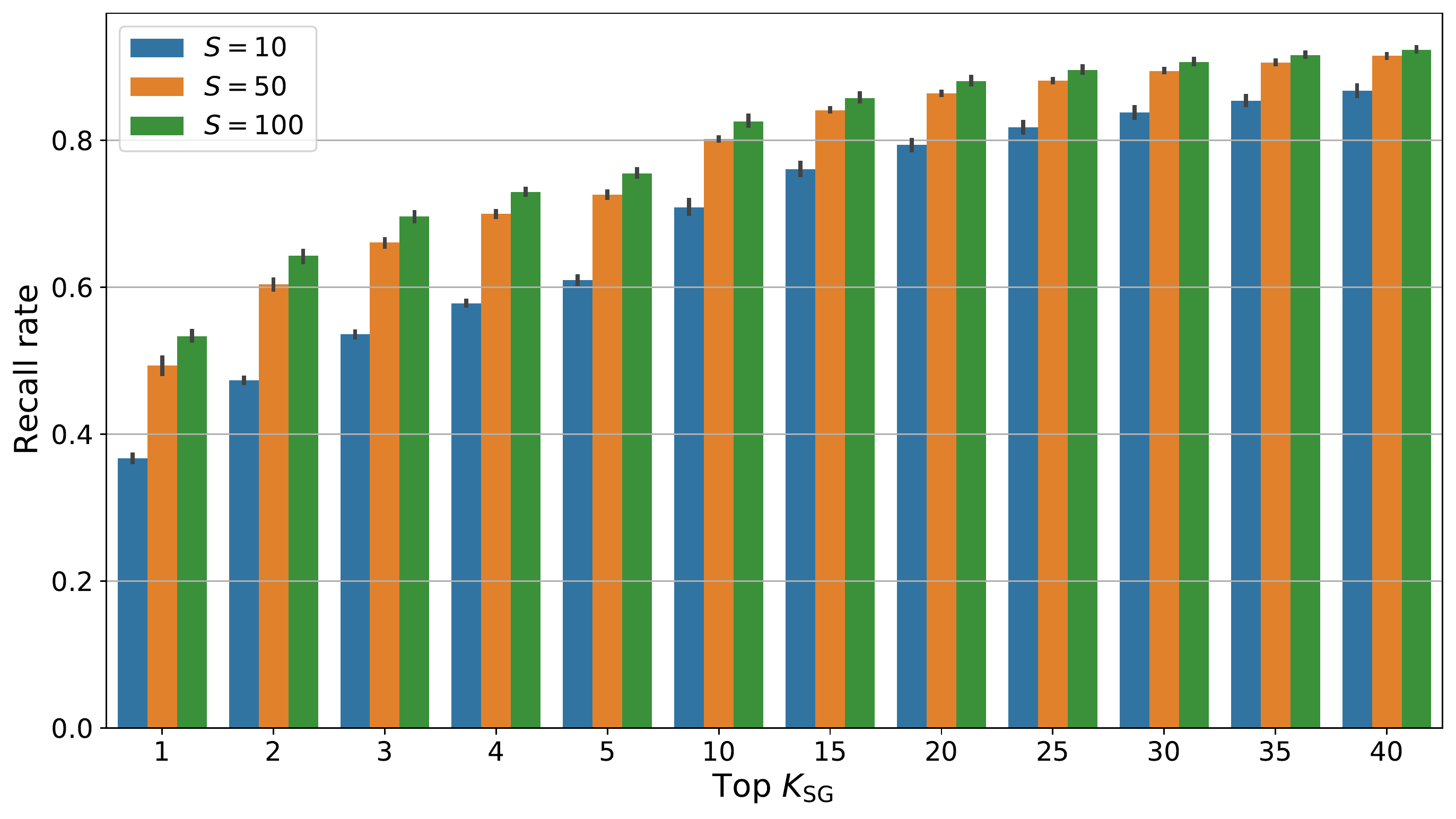}
\caption{Recall rates of space-group prediction with different upper bounds $S$ on the number of samples with the same composition ratio in the training set ($S \in \{10, 50, 100\}$).}
\label{fig:S3_recall_sensitivity}
% \addtocounter{figure}{1}
\end{figure}

The possibility of significant bias in the patterns of composition ratios in the dataset is a concern because this could improve the prediction performance for test instances that have many identical composition ratios in the training set. To address this concern, we set an upper bound $S$ on the number of samples with the same composition ratio in the training set, and then evaluated the sensitivity of the prediction accuracy of the trained models to $S$ by varying $S \in \{10, 50, 100\}$. The results in Figure \ref{fig:S3_recall_sensitivity} confirm that the prediction performance did not vary significantly with $S$.

\begin{figure}[!ht]
\centering
\includegraphics[width=0.95\linewidth]{suppl/recall_rates/1.pdf}
\captionsetup{labelformat=empty}
\caption[]{}
\end{figure}
\begin{figure}[ht]\ContinuedFloat
\centering
\includegraphics[width=0.95\linewidth]{suppl/recall_rates/2.pdf}
\caption{Recall rate of the top 30 predictions and number of training instances for each space group. Training and testing were repeated independently 100 times. The color bars show the average recall rate, with the error bars representing the standard deviation. The red dots show the average number of training instances.}
\label{fig:S2_recall_each_spg}
% \addtocounter{figure}{1}
\end{figure}

\section*{Prediction of unit-cell volume}
\addcontentsline{toc}{section}{Prediction of unit-cell volume}

Using 33,040 stable crystal structures from the Materials Project database and their unit-cell volumes, excluding the 120 benchmark crystal structures, we constructed a fully connected neural network to predict the unit-cell volume for any composition. The samples were split into training and test sets in a 4:1 ratio. Fivefold cross-validation was performed with the training set to identify the hyperparameters with the highest average prediction accuracy. Specifically, the hyperparameters were selected by Bayesian optimization using the Optuna Python library (the number of trials was set to 200) from the candidate set. The number of layers $\in \{2,3,4\}$, dropout ratio $\in \{0,0.1,0.2\}$, and number of neurons in each layer ($N \times \kappa$, where $N$ denotes the number of neurons in the previous layer and $\kappa \in [0.8, 0.95]$) were optimized. Finally, we used the test set to calculate the predictive performance of the trained neural network with the hyperparameters selected by Optuna. The performance metrics were averaged over 100 bootstrap sets. The mean absolute error (MAE), root mean square error (RMSE), and coefficient of determination ($\mathrm{R}^2$) were $53.048 \pm 3.177$ \AA, $90.362 \pm 6.007$ \AA, and $0.973 \pm 0.004$, respectively. 

\section*{VASP calculation inputs}
\addcontentsline{toc}{section}{VASP calculation inputs}

For DFT calculations, the VASP ``INCAR'' files were generated through a Python script by leveraging the \texttt{MPStaticSet} and \texttt{MPRelaxSet} presets of pymatgen \cite{ong2013PythonMaterials} with some modifications. The Python script was used to iterate through a dataset of structures by implementing a series of relaxations. Each relaxation step was uniquely parameterized to optimize the computational process and accuracy of the results.

\begin{itemize}
    \item[] \textbf{Step 1: Initial relaxation}
    \begin{itemize}
        \item Utilize MPRelaxSet with lower accuracy for coarse relaxation.
        \item Modified parameters include:
        \begin{itemize}
            \item \texttt{ALGO=Fast} for quick electronic minimization.
            \item \texttt{EDIFF=1e-2}, \texttt{EDIFFG=1e-1} for looser convergence criteria.
            \item \texttt{ISIF=4} to relax ion positions, cell shape, and volume.
            \item \texttt{PREC=LOW}, \texttt{POTIM=0.02}, and \texttt{NSW=90} for ionic relaxation.
        \end{itemize}
    \end{itemize}

    \item[] \textbf{Step 2: Intermediate relaxation}
    \begin{itemize}
        \item Continue with MPRelaxSet for further relaxation.
        \item Parameters include \texttt{EDIFF=1e-3} and \texttt{EDIFFG=1e-2} for tighter convergence and \texttt{IBRION=1} for conjugate gradient algorithm.
        \item \texttt{PREC=Normal} and \texttt{POTIM=0.3} set for moderate accuracy.
    \end{itemize}

    \item[] \textbf{Step 3: Further relaxation}
    \begin{itemize}
        \item Utilize MPRelaxSet to balance accuracy and computational efficiency.
        \item Include \texttt{IBRION=2}, \texttt{ISIF=3}, and \texttt{SIGMA=0.1} for consistent relaxation and electronic convergence.
    \end{itemize}

    \item[] \textbf{Step 4: Pre-static calculation}
    \begin{itemize}
        \item Employ MPRelaxSet with a focus on higher accuracy.
        \item Key parameters include \texttt{EDIFF=1e-4} and \texttt{EDIFFG=1e-3} for tight convergence and \texttt{PREC=Accurate} for enhanced precision.
    \end{itemize}

    \item[] \textbf{Step 5: Static calculation}
    \begin{itemize}
        \item Use MPStaticSet for final static computations.
        \item Parameters included \texttt{ALGO=Fast} and \texttt{EDIFF=1e-4} for electronic minimization and \texttt{IBRION=-1} and \texttt{ISMEAR=-5} tailored for static calculations.
        \item \texttt{PREC=Accurate}, \texttt{SIGMA=0.05}, and \texttt{NSW=0} were used to ensure high precision and no ionic relaxation.
    \end{itemize}
\end{itemize}

For all calculations, we used a consistent plane-wave cutoff energy \texttt{ENCUT=520} eV. This ensured sufficient accuracy across all types of calculations, from initial relaxation to final static analysis, which is crucial for achieving reliable and consistent simulation results.

The above VASP settings were strategically designed to ensure the stability and applicability of the structural optimization process, while also considering computational efficiency. For simpler compounds such as those containing C and Si, which contain fewer atoms, we optimized the computational speed by strategically adjusting parameters such as ``NELM,'' ``NSW,'' and ``EDIFF.'' These adjustments improved the calculation speed without compromising the integrity or reliability of the simulation results.

\section*{USPEX calculation inputs}
\addcontentsline{toc}{section}{USPEX calculation inputs}

We developed a Python script for the systematic generation of input files for USPEX calculations by adhering closely to the official USPEX guidelines. A carefully prepared template file served as the foundation for this process. This file contained placeholders that represented essential parameters for USPEX calculations, including aspects of the evolutionary algorithm, population settings, and variation operators.

For each composition and its corresponding ground-truth structure, the Python script performed a detailed extraction of the element types, element quantities, and space-group information. These extracted values were then used as substitutes for the respective placeholders in the template, namely \texttt{atomType}, \texttt{numSpecies}, and \texttt{symmetries}.

The parameter \texttt{populationSize} was calculated by aggregating the quantities of each constituent element within a structure and rounding to the nearest ten. An upper limit of 60 was imposed to maintain computational efficiency. In parallel, the \texttt{stopCrit} parameter was derived in a congruent manner by employing the total atom count, rounded up to the nearest ten, to define the termination criterion for the evolutionary exploration.

The \texttt{fracAtomsMut} parameter was set to 0.20 for compositions comprising a single element, reflecting a higher mutation rate, which is suitable for systems of lesser complexity. Conversely, for compositions featuring multiple elements, this parameter was reduced to 0.10 to consider the greater complexity and potential variations in the energy landscape of such systems.

Lastly, the \texttt{keepBestHM} parameter was calculated as 15\% of \texttt{populationSize}, rounded to the nearest whole number. This approach was strategically designed to strike a balance between preserving the most promising structures from the current generation and fostering the exploration of novel configurations in future generations.

This systematic and rigorous parameterization process ensured that each generated ``INPUT.txt'' file was specifically tailored to the unique attributes of each individual structure, thereby optimizing the efficacy and precision of the evolutionary algorithm within USPEX. To exemplify this process, we present the ``INPUT.txt'' file for the composition \ce{Ag32Ge4S24} as a demonstrative case:

\begin{displayquote}
**************************\\
TYPE OF RUN AND SYSTEM\\
***************************\\
USPEX : calculationMethod\\
300   : calculationType\\
1     : AutoFrac\\

\% optType\\
1\\
\% EndOptType\\

\% atomType\\
Ag Ge S\\
\% EndAtomType\\

\% numSpecies\\
32 4 24\\
\% EndNumSpecies

\% symmetries\\ 
33\\
\% EndSymmetries\\

*************\\
POPULATION\\
**************\\
60  : populationSize\\
60  : initialPopSize\\
100   : numGenerations\\
0.00  : reoptOld\\
0.60  : bestFrac\\
9  : keepBestHM\\
60  : stopCrit\\

***********************\\
VARIATION OPERATORS\\
************************\\
0.40  : fracGene\\
0.20  : fracRand\\
0.00  : fracRotMut\\
0.20  : fracTopRand\\
0.10  : fracAtomsMut\\
0.00  : fracLatMut\\
0.10  : fracPerm\\

*************************************\\
DETAILS OF AB INITIO CALCULATIONS\\
**************************************\\
\% abinitioCode\\
1 1 1 1 1\\
\% ENDabinit\\

\% KresolStart\\
0.14 0.12 0.11 0.09 0.06\\
\% Kresolend
\end{displayquote}

\section*{DBSCAN clustering}
\addcontentsline{toc}{section}{DBSCAN clustering}

DBSCAN is a clustering method that forms clusters based on the density of data points. The procedure is summarized as follows.

\begin{enumerate}
  \item \textbf{Definition of $\varepsilon$-neighborhood}:
    Define the $\varepsilon$-neighborhood for each data point using a given distance threshold $\varepsilon$ as a parameter. This neighborhood is the set of data points within a distance $\varepsilon$ from a given data point.

  \item \textbf{Identification of core points}:
    If the $\varepsilon$-neighborhood of a data point contains at least $N_{\min}$ data points, that data point is considered a core point.

  \item \textbf{Direct density reachability}:
    Data point $i$ is considered directly density-reachable from data point $j$ if there is a chain of continuous core points from $i$ to $j$. This indicates that the core points are densely connected, implying that they belong to the same cluster.

  \item \textbf{Cluster formation}:
  Core points that are directly density-reachable are assigned to the same cluster. In this manner, the entire dataset is divided into clusters.

  \item \textbf{Identification of noise points}:
    Data points that are not directly density-reachable from core points are considered noise points. These points do not belong to any cluster and are isolated.
\end{enumerate}

A key advantage of DBSCAN is that it does not require the number of clusters to be specified beforehand. It can detect clusters of arbitrary shapes and identify outliers (noise points). However, selecting appropriate values for $\varepsilon$ and $N_{\min}$ can be challenging, particularly in datasets with varying densities. In this study, we set $\varepsilon = 9$ and $N_{\min} = 10$.

\section*{Energy comparison with ground-truth structures}
\addcontentsline{toc}{section}{Energy comparison with ground-truth structures}

To gain deeper insight into cases where structure prediction was unsuccessful, we investigated the total energy differences between all predicted relaxed structures and their corresponding ground-truth structures. These calculations utilized the VASP input parameters described above. The analysis revealed several notable facts:

\begin{itemize}
    \item Although the ground-truth structures typically exhibited the lowest energies, several exceptions were observed, such as for \ce{AlH12(ClO2)3}, \ce{VCl5}, and the USPEX-proposed structure for \ce{Pm2NiIr}. In these instances, the predicted structures exhibited lower energies than those found in the Materials Project, revealing three previously unknown ground states.
    
    \item For \ce{Zr4O} and \ce{CsPbI3}, the predicted structures had slightly higher energies than their ground-truth counterparts.
\end{itemize}

\begin{figure}[!ht]
\centering
\includegraphics[width=0.95\linewidth]{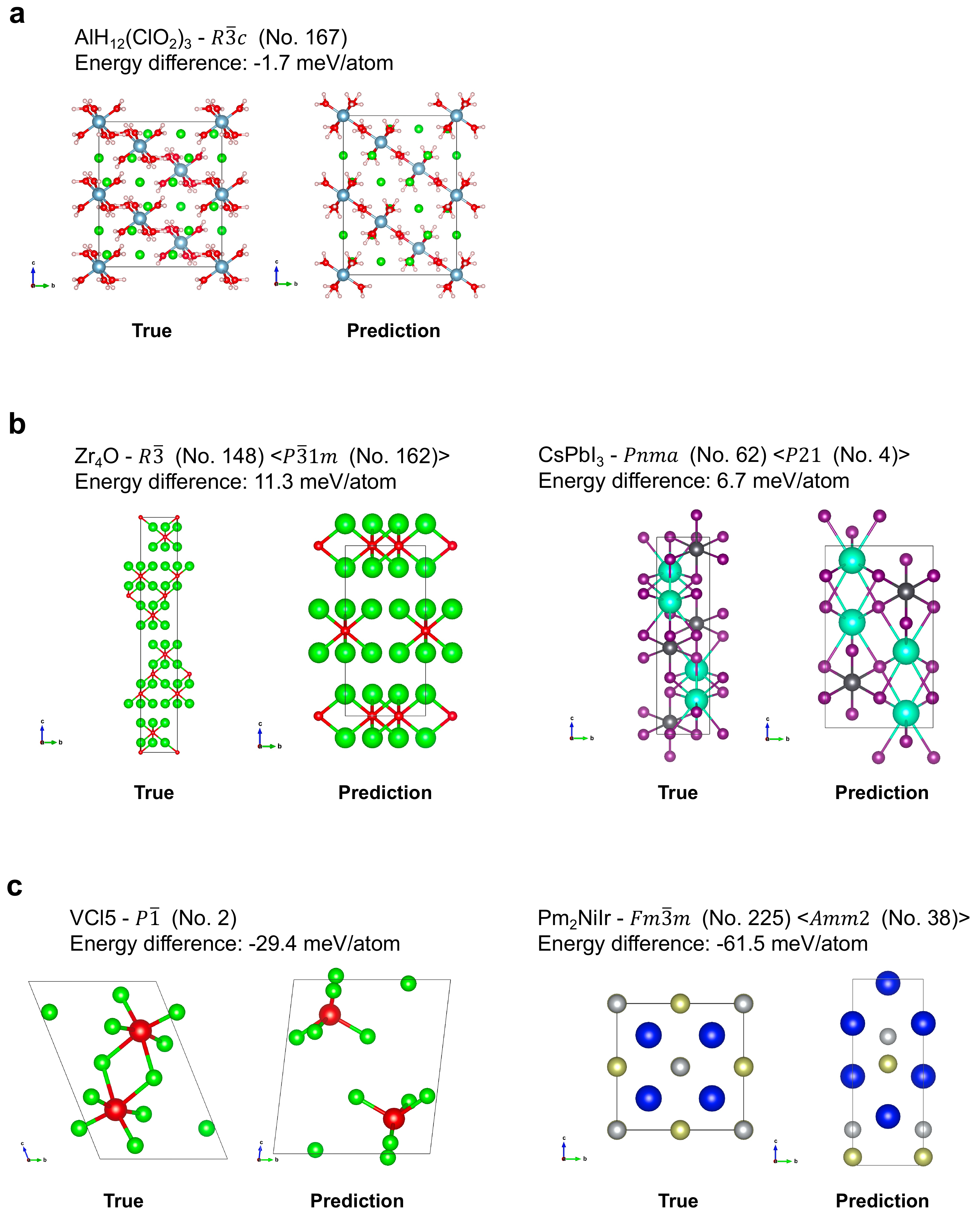}
\caption{Comparison of proposed structures and their corresponding ground-truth structures obtained from Materials Project. The negative sign of the energy indicates the proposed structure has a lower energy compared to the ground-truth structure. If the space groups of the proposed and ground-truth structures differ, the space group of the proposed structure is indicated in angle brackets.}
\label{fig:en_compare}
\end{figure}

In the case of \ce{AlH12(ClO2)3} (Figure~\ref{fig:en_compare}(a)), the predicted structures closely resembled the ground-truth structures; however, they exhibited higher symmetry in the orientation of the \ce{-OH} sites. This resulted in a 2 \AA\ extension along the $c$ axis, potentially contributing to the observed energy difference of $-1.7$ meV/atom.

For \ce{Zr4O} and \ce{CsPbI3} (Figure~\ref{fig:en_compare}(b)), the lowest energy predicted structures were similar to the ground-truth structures but had slightly higher energies, with differences of +11.3 and +6.7 meV/atom, respectively. Consequently, these cases were categorized as failed predictions.

In the cases of \ce{VCl5} and USPEX-proposed \ce{Pm2NiIr} (Figure~\ref{fig:en_compare}(c)), energy differences of $-29.4$ and $-61.5$ meV/atom, respectively, were observed compared to their ground-truth structures. These cases were classified as previously unknown ground states.

%% Description of CSPML model added by kusaba.
\section*{Details of CSPML}
\addcontentsline{toc}{section}{{Details of CSPML}}
Here, we describe the training details and model architecture of CSPML \cite{kusaba2022CrystalStructure}. The objective of CSPML is to obtain a predictor that, given any two compositions, determines whether their stable structures are identical. With this predictor, compositions sharing the same stable structure as a given query composition can be efficiently screened from a crystal structure database. The crystal structure is predicted by applying element substitutions to the stable structure set predicted to be identical, and performing structure relaxation by first-principles calculations.

A training set was sampled from the 126,210 crystals in the Materials Project database, excluding the 120 benchmark crystals. 
The dataset exhibited significant bias toward specific composition ratios. To remove this bias, we trained and evaluated the CSPML in a slightly different manner to that in the previous study \cite{kusaba2022CrystalStructure}. We extracted all stable structures (defined as energy above hull = 0) from the Materials Project database, resulting in 33,064 stable structures, each with a unique chemical composition. These stable structures were randomly divided into training, validation, and test sets in a 3:1:1 ratio. To obtain a model ensemble, we repeated the data splitting independently five times using different random seeds. To reduce bias in the dataset toward certain composition ratios---for instance, 3892 structures had a composition ratio of 2:1:1---the maximum number of samples with the same composition ratio was set to 1\% of the total dataset size (e.g., 100 for a dataset size of 10,000). Specifically, data with composition ratios exceeding this limit underwent downsampling. This downsampling process was applied to each of the training, validation, and test sets. 

To obtain class labels, we calculated the structural dissimilarity of all pairs of crystal structures with the same composition ratio, as described in the Methods section in the main text. Pairs with a structural dissimilarity $\tau$ smaller than 0.2 were categorized as similar, while those with dissimilarity exceeding this threshold were classified as dissimilar. Because the number of similar pairs was much smaller than the number of dissimilar pairs, downsampling was performed until the number of dissimilar pairs equaled the number of similar pairs.

\begin{figure}[!ht]
\centering
\includegraphics[width=0.95\linewidth]{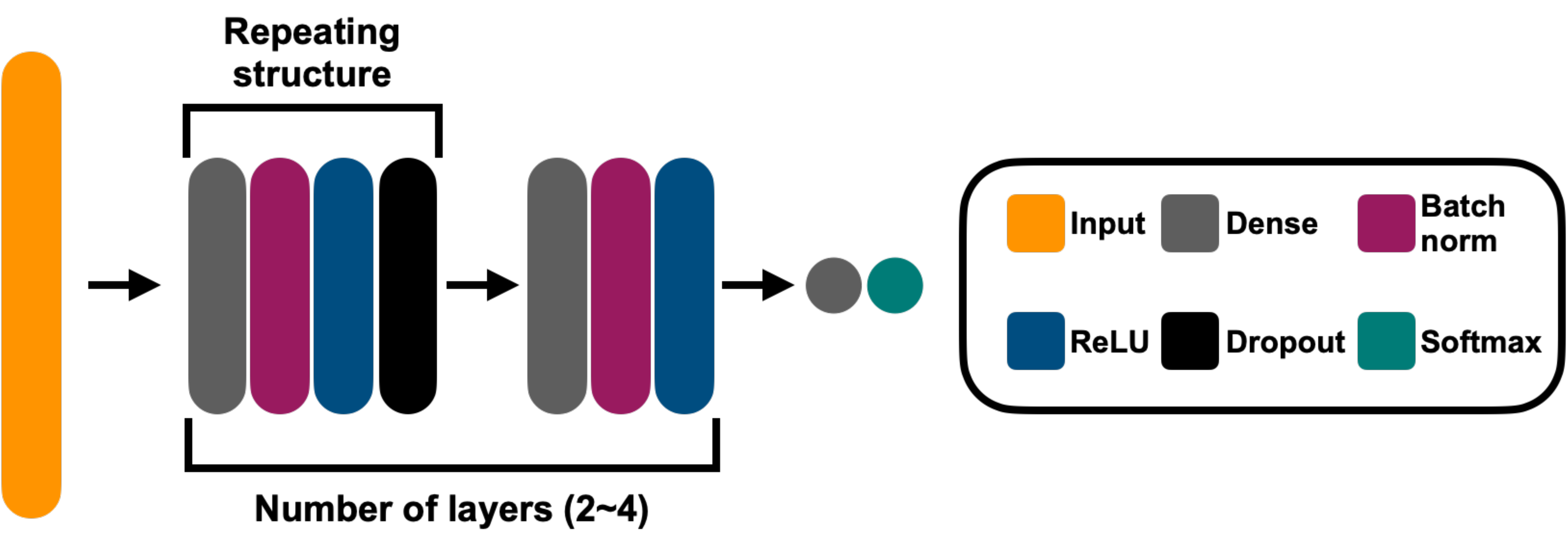}
\caption{Architecture of fully connected neural network for CSPML. The input variable consists of two chemical compositions, and the output variable represents whether or not their stable structures are identical.}
\label{CSPML_model}
% \addtocounter{figure}{1}
\end{figure}

The CSPML architecture is depicted in Figure \ref{CSPML_model}. The input variable involves a pair of two compositions, $X_i$ and $X_j$. Features of each composition are encoded into 290-dimensional descriptor vectors $\phi$ using XenonPy. CSPML then defines a mapping from the absolute distance $|\phi(X_i) - \phi(X_j)|$ to the binary class label indicating whether or not their stable structures are identical. Specifically, the model adopts a fully connected neural network architecture comprising densely connected layers (Dense), rectified linear units (ReLU), dropout layers (Dropout), batch normalization layers (Batch norm), and a softmax function as the final activation function for binary classification. The network consists of one or more repeating structures with a dropout layer and an output layer without dropout. All intermediate layers have an identical number of units.

The hyperparameters of the models were selected using Bayesian optimization with the Optuna Python library by employing the validation set (with 50 trials). The search space included the number of layers $\in \{2,3,4\}$, dropout rate $\in \{0,0.1,0.2\}$, number of neurons in each layer $\in \{200,400,600,800,1000\}$, batch size $\in \{1024, 2048\}$, and patience for early stopping $\in \{50,75,100\}$. Each model was trained until the validation error converged (with a patience hyperparameter varying from 50 to 100 epochs) or until 1000 epochs were reached. The Adam optimizer ($\beta_1 = 0.9$, $\beta_2 = 0.999$) was employed for back-propagating gradients. Training was performed using the TensorFlow-macOS v2.9.0 library, with the TensorFlow-metal v0.5.1 plug-in utilized for GPU calculations (Apple M1 Max, 32 GPU cores). To reduce the computational cost, we used only the first training and validation datasets in the Bayesian optimization.
The resulting hyperparameters were as follows: number of layers, 3; dropout rate, 0.2; number of neurons in each layer, 400; batch size, 1024; and patience, 50. With these hyperparameters, we trained the models on all five training and validation datasets. The data and code utilized to train the CSPML are available on GitHub \cite{CSPMLgithub}.

Performance metrics such as the mean average precision (MAP), mean accuracy (MACC), average precision (AP), and accuracy (ACC) with respect to the test sets are presented in Table \ref{table:CSPMLpredictions}. MAP and MACC were calculated by averaging the AP and ACC calculated for each composition ratio. The performance metrics were averaged over the five trials and listed as the mean $\pm$ standard deviation. These results demonstrate that the models can accurately predict structural similarity from chemical composition pairs with identical composition ratios.

An ensemble of the five models, denoted as $f_1, \ldots, f_5$, was leveraged to calculate the predicted class label. The probability of being classified into similar pairs was determined by $\hat{f}(|\phi(X_i)-\phi(X_j)|) = \frac{1}{5}\sum_{b=1}^{5} f_b(\phi(X_i)-\phi(X_j)|)$. For a given query composition $X$, a set of candidate templates was selected from the 33,064 stable structures, such that their predicted class labels were classified as being identical to $X$. The elements of $X$ were substituted into each template structure, and then DFT structure relaxation was performed to obtain a predicted crystal structure. The computational environment and procedure of the DFT calculations were the same as those in the shotgun CSP workflow.

For each query composition of the 90 benchmark crystals, the top ten structures were predicted. However, the CSPML could not propose any templates for eight cases. For \ce{NaCaAlPHO5F2}, \ce{K5Ag2(AsSe3)3}, \ce{Na(WO3)9}, \ce{Li6V3P8O29}, and \ce{Mg3Si2H4O9}, none of the candidates shared the same composition ratio in the pool of 33,064 candidates. For \ce{MgB7}, \ce{Ba2CaSi4(BO7)2}, and \ce{Y4Si5Ir9}, none of the candidates exhibited class probabilities greater than 0.5. The results of CSP using CSPML are summarized in Table~\ref{tab:CSPML_results}. 

\begin{table*}[ht]
\caption{Prediction performance of CSPML: average precision (MAP), mean accuracy (MACC), average precision (AP), and accuracy (ACC). Values are shown as the mean $\pm$ standard deviation.}
\centering
\begin{tabular}{c c c c}
\hline\hline
MAP & MACC & AP & ACC\\ [0.5ex] 
\hline
$0.941 \pm{0.007}$ & $0.890 \pm{0.007}$ & $0.913 \pm{0.009}$  & $0.864 \pm{0.011}$  \\
[1ex]
\hline
\end{tabular}
\label{table:CSPMLpredictions}
\end{table*}

\clearpage

\renewcommand\theadalign{cc}
\renewcommand\theadfont{\bfseries}
\renewcommand\theadgape{\Gape[3pt]}
\renewcommand\cellgape{\Gape[4pt]}

\begin{longtable}[!t]{r|cccc}
\caption{Prediction performance of CSPML for the 90 crystals comprising Datasets I and II. The top ten virtual structures were proposed as the final candidates. The second column indicates the number of atoms in the primitive unit cells. In the fourth column, the symbols $\checkmark$ and $\times$ indicate prediction success and failure, respectively, and  dashes ($-$) indicate cases for which there was no template for element substitution. The symbols in parentheses indicate whether a similar template structure ($\tau \leq 0.2$) was found.}

\label{tab:CSPML_results}\\
\wcline{1-5}
    \thead{Composition} & \thead{Number of atoms} & \thead{Space group} & \thead{CSP using\\CSPML} \\
\cline{1-5}
\endfirsthead 

\multicolumn{5}{l}{Table \ref{tab:CSPML_results} continued}\\
\wcline{1-5}
    \thead{Composition} & \thead{Number of atoms} & \thead{Space group} & \thead{CSP using\\CSPML} \\
\cline{1-5}
\endhead

\wcline{1-5} 
\endfoot

\endlastfoot
    \ce{C}              & 4 & $R\bar3m$      & $\times$ ($\checkmark$) \\ \hline
    \ce{Si}             & 2 & $Fd\bar3m$     & $\checkmark$ ($\checkmark$) \\ \hline
    \ce{GaAs}           & 2 & $F\bar43m$     & $\checkmark$ ($\checkmark$) \\ \hline
    \ce{ZnO}            & 4 & $P6_3mc$       & $\times$ ($\checkmark$) \\ \hline
    \ce{BN}             & 4 & $P6_3/mmc$     & $\checkmark$ ($\checkmark$) \\ \hline
    \ce{LiCoO2}         & 16 & $R\bar3m$     & $\checkmark$ ($\checkmark$) \\ \hline
    \ce{Bi2Te3}         & 5 & $R\bar3m$      & $\times$ ($\checkmark$) \\ \hline
    \ce{Ba(FeAs)2}      & 5 & $I4/mmm$       & $\checkmark$ ($\checkmark$) \\ \hline
    \ce{SiO2}           & 6 & $I\bar42d$     & $\checkmark$ ($\checkmark$) \\ \hline
    \ce{VO2}            & 6 & $P4_2/mnm$     & $\checkmark$ ($\checkmark$) \\ \hline
    \ce{La2CuO4}        & 7 & $I4/mmm$       & $\checkmark$ ($\checkmark$) \\ \hline
    \ce{LiPF6}          & 8 & $R\bar3$       & $\checkmark$ ($\checkmark$) \\ \hline
    \ce{Al2O3}          & 10 & $R\bar3c$     & $\checkmark$ ($\checkmark$) \\ \hline
    \ce{SrTiO3}         & 10 & $I4/mcm$      & $\times$ ($\checkmark$) \\ \hline
    \ce{CaCO3}          & 10 & $R\bar3c$     & $\checkmark$ ($\checkmark$)  \\ \hline
    \ce{TiO2}           & 12 & $C2/m$        & $\times$ ($\checkmark$) \\ \hline
    \ce{ZrO2}           & 12 & $P2_1/c$      & $\checkmark$ ($\checkmark$) \\ \hline
    \ce{ZrTe5}          & 12 & $Cmcm$        & $\checkmark$ ($\checkmark$) \\ \hline
    \ce{V2O5}           & 14 & $Pmmn$        & $-$ \\ \hline
    \ce{Si3N4}          & 14 & $P6_3/m$      & $\times$ ($\checkmark$) \\ \hline
    \ce{Fe3O4}          & 14 & $Fd\bar3m$    & $\checkmark$ ($\checkmark$) \\ \hline
    \ce{Mn(FeO2)2}      & 14 & $Fd\bar3m$    & $\checkmark$ ($\checkmark$) \\ \hline
    \ce{ZnSb}           & 16 & $Pbca$        & $\times$ ($\checkmark$) \\ \hline
    \ce{CoSb3}          & 16 & $Im\bar3$     & $\checkmark$ ($\checkmark$) \\ \hline
    \ce{LiBF4}          & 18 & $P3_121$      & $\times$ ($\checkmark$) \\ \hline
    \ce{Y2Co17}         & 19 & $R\bar3m$     & $\checkmark$ ($\checkmark$) \\ \hline
    \ce{GeH4}           & 20 & $P2_12_12_1$  & $\times$ ($\checkmark$) \\ \hline
    \ce{CsPbI3}         & 20 & $Pnma$        & $\checkmark$ ($\checkmark$) \\ \hline
    \ce{NaCaAlPHO5F2}   & 24 & $P2_1/m$      & $-$ \\ \hline
    \ce{LiFePO4}        & 28 & $Pnma$        & $\checkmark$ ($\checkmark$) \\ \hline
    \ce{Cu12Sb4S13}     & 29 & $I\bar43m$    & $\checkmark$ ($\checkmark$) \\ \hline
    \ce{MgB7}           & 32 & $Imma$        & $-$ \\ \hline
    \ce{Li3PS4}         & 32 & $Pnma$        & $\times$ ($\checkmark$) \\ \hline
    \ce{Cd3As2}         & 80 & $I4_1/acd$    & $\times$ ($\checkmark$) \\ \hline
    \ce{Li4Ti5O12}      & 42 & $C2/c$        & $\times$ ($\checkmark$) \\ \hline
    \ce{Ba2CaSi4(BO7)2} & 46 & $I\bar42m$    & $-$ \\ \hline
    \ce{Ag8GeS6}        & 60 & $Pna2_1$      & $\checkmark$ ($\checkmark$) \\ \hline
    \ce{Nd2Fe14B}       & 68 & $P4_2/mnm$    & $\times$ ($\checkmark$) \\ \hline
    \ce{Y3Al5O12}       & 80 & $Ia\bar3d$    & $-$ \\ \hline
    \ce{Ca14MnSb11}     & 104 & $I4_1/acd$   & $\checkmark$ ($\checkmark$) \\ \hline
    \ce{CsCl}           & 2 & $Fm\bar3m$      & $\checkmark$ ($\checkmark$) \\ \hline
    \ce{MnAl}           & 2 & $P4/mmm$        & $\checkmark$ ($\checkmark$) \\ \hline
    \ce{HoHSe}          & 3 & $P\bar6m2$      & $\checkmark$ ($\checkmark$) \\ \hline
    \ce{ErCdRh2}        & 4 & $Fm\bar3m$      & $\checkmark$ ($\checkmark$) \\ \hline
    \ce{Eu2MgTl}        & 4 & $Fm\bar3m$      & $\checkmark$ ($\checkmark$) \\ \hline
    \ce{Pm2NiIr}        & 4 & $Fm\bar3m$      & $\checkmark$ ($\checkmark$) \\ \hline
    \ce{VPt3}           & 4 & $I4/mmm$        & $\checkmark$ ($\checkmark$) \\ \hline
    \ce{Gd(SiOs)2}      & 5 & $I4/mmm$        & $\checkmark$ ($\checkmark$) \\ \hline
    \ce{LaAl3Au}        & 5 & $I4mm$          & $\checkmark$ ($\checkmark$) \\ \hline
    \ce{U2SbN2}         & 5 & $I4/mmm$        & $\checkmark$ ($\checkmark$) \\ \hline
    \ce{MnGa(CuSe2)2}   & 8 & $I\bar4$        & $\checkmark$ ($\checkmark$) \\ \hline
    \ce{SmZnPd}         & 9 & $P\bar62m$      & $\checkmark$ ($\checkmark$) \\ \hline
    \ce{Sn(TePd3)2}     & 9 & $I4mm$          & $\checkmark$ ($\checkmark$) \\ \hline
    \ce{V5S4}           & 9 & $I4/m$          & $\checkmark$ ($\checkmark$) \\ \hline
    \ce{Cs3InF6}        & 10 & $Fm\bar3m$     & $\checkmark$ ($\checkmark$) \\ \hline
    \ce{Eu(CuSb)2}      & 10 & $P4/nmm$       & $\checkmark$ ($\checkmark$) \\ \hline
    \ce{Rb2TlAgCl6}     & 10 & $Fm\bar3m$     & $\checkmark$ ($\checkmark$) \\ \hline
    \ce{Ca3Ni7B2}       & 12 & $R\bar3m$      & $\checkmark$ ($\checkmark$) \\ \hline
    \ce{DyPO4}          & 12 & $I4_1/amd$     & $\checkmark$ ($\checkmark$) \\ \hline
    \ce{LaSiIr}         & 12 & $P2_13$        & $\checkmark$ ($\checkmark$) \\ \hline
    \ce{SmVO4}          & 12 & $I4_1/amd$     & $\checkmark$ ($\checkmark$) \\ \hline
    \ce{VCl5}           & 12 & $P\bar1$       & $\checkmark$ ($\checkmark$) \\ \hline
    \ce{YbP5}           & 12 & $P2_1/m$       & $\checkmark$ ($\checkmark$) \\ \hline
    \ce{Eu(Al2Cu)4}     & 13 & $I4/mmm$       & $\checkmark$ ($\checkmark$) \\ \hline
    \ce{Zr4O}           & 15 & $R\bar3$       & $\times$ ($\checkmark$)         \\ \hline
    \ce{Ba3Ta2NiO9}     & 15 & $P\bar3m1$     & $\checkmark$ ($\checkmark$)         \\ \hline
    \ce{K2Ni3S4}        & 18 & $Fddd$         & $\checkmark$ ($\checkmark$) \\ \hline
    \ce{Sr(ClO3)2}      & 18 & $Fdd2$         & $\checkmark$ ($\checkmark$) \\ \hline
    \ce{LiSm2IrO6}      & 20 & $P2_1/c$       & $\checkmark$ ($\checkmark$) \\ \hline
    \ce{Pr2ZnPtO6}      & 20 & $P2_1/c$       & $\checkmark$ ($\checkmark$) \\ \hline
    \ce{Sc2Mn12P7}      & 21 & $P\bar6$       & $\checkmark$ ($\checkmark$) \\ \hline
    \ce{LaSi2Ni9}       & 24 & $I4_1$/amd     & $\checkmark$ ($\checkmark$) \\ \hline
    \ce{CeCu5Sn}        & 28 & $Pnma$         & $\checkmark$ ($\checkmark$) \\ \hline
    \ce{LiP(HO2)2}      & 32 & $Pna2_1$       & $-$     \\ \hline
    \ce{Mg3Si2H4O9}     & 36 & $P6_3cm$       & $-$         \\ \hline
    \ce{Y4Si5Ir9}       & 36 & $P6_3/mmc$     & $-$                         \\ \hline
    \ce{Na(WO3)9}       & 37 & $R\bar3$       & $-$                         \\ \hline
    \ce{Sm6Ni20As13}    & 39 & $P\bar6$       & $\checkmark$ ($\checkmark$) \\ \hline
    \ce{BaCaGaF7}       & 40 & $P2/c$         & $\checkmark$ ($\checkmark$) \\ \hline
    \ce{Tm11Sn10}       & 42 & $I4/mmm$       & $-$ \\ \hline
    \ce{AlH12(ClO2)3}   & 44 & $R\bar3c$      & $\checkmark$ ($\checkmark$) \\ \hline
    \ce{K2ZrSi2O7}      & 48 & $P2_1/c$       & $\times$ ($\checkmark$)     \\ \hline
    \ce{LiZr2(PO4)3}    & 72 & $P2_1/c$       & $\checkmark$ ($\checkmark$) \\ \hline
    \ce{K5Ag2(AsSe3)3}  & 76 & $Pnma$         & $-$                         \\ \hline
    \ce{Be17Ru3}        & 80 & $Im\bar3$      & $-$ \\ \hline
    \ce{Cu3P8(S2Cl)3}   & 80 & $Pnma$         & $\checkmark$ ($\checkmark$) \\ \hline
    \ce{Al2CoO4}        & 84 & $P3m1$         & $\times$ ($\checkmark$) \\ \hline
    \ce{Li6V3P8O29}     & 92 & $P1$           & $-$     \\ \hline
    \ce{ReBi3O8}        & 96 & $P2_13$        & $\times$ ($\checkmark$) \\ \hline
    \ce{Na5FeP2(O4F)2}  & 288 & $Pbca$        & $\times$ ($\checkmark$)               \\ \hline
\wcline{1-5}
\multicolumn{1}{r}{\textbf{Overall}}        & & & $59 / 90 = 65.6\%$ \\

\bottomrule
\end{longtable}

\clearpage

\begin{table}[ht!]
\centering
\caption{
Results of Wyckoff-position-generator-based CSP (ShotgunCSP-GW) and USPEX for 25 crystals from Datasets I and II. See the caption of Table 1 in the main text for details.}
\begin{tabular}{r|ccccc}
\wcline{1-6}
    \thead{Composition} & \thead{Number of atoms} & \thead{Space group} & \thead{Dataset} & \thead{ShotgunCSP-GW} & \thead{USPEX} \\
\cline{1-6}
    \ce{C}              & 4 & $R\bar3m$     & I & $\checkmark$ ($\checkmark$ \slash\ $\checkmark$)    & $\checkmark$ \\ \hline
    \ce{GaAs}           & 2 & $F\bar43m$    & I & $\checkmark$ ($\checkmark$ \slash\ $\checkmark$)    & $\checkmark$ \\ \hline
    \ce{ZnO}            & 4 & $P6_3mc$      & I & $\checkmark$ ($\checkmark$ \slash\ $\checkmark$)    & $\checkmark$ \\ \hline
    \ce{BN}             & 4 & $P6_3/mmc$    & I & $\checkmark$ ($\checkmark$ \slash\ $\checkmark$)    & $\checkmark$ \\ \hline
    \ce{LiCoO2}         & 16 & $R\bar3m$    & I & $\checkmark$ ($\checkmark$ \slash\ $\checkmark$)    & $\checkmark$ \\ \hline
    \ce{Bi2Te3}         & 5 & $R\bar3m$     & I & $\checkmark$ ($\checkmark$ \slash\ $\checkmark$)    & $\checkmark$ \\ \hline
    \ce{Ba(FeAs)2}      & 5 & $I4/mmm$      & I & $\checkmark$ ($\checkmark$ \slash\ $\checkmark$)    & $\checkmark$ \\ \hline
    \ce{La2CuO4}        & 7 & $I4/mmm$      & I & $\times$ ($\checkmark$ \slash\ $\checkmark$)        & $\checkmark$ \\ \hline
    \ce{Al2O3}          & 10 & $R\bar3c$    & I & $\checkmark$ ($\checkmark$ \slash\ $\checkmark$)    & $\checkmark$ \\ \hline
    \ce{SrTiO3}         & 10 & $I4/mcm$     & I & $\checkmark$ ($\checkmark$ \slash\ $\checkmark$)    & $\checkmark$ \\ \hline
    \ce{CaCO3}          & 10 & $R\bar3c$    & I & $\checkmark$ ($\checkmark$ \slash\ $\checkmark$)    & $\checkmark$ \\ \hline
    \ce{Fe3O4}          & 14 & $Fd\bar3m$   & I & $\checkmark$ ($\checkmark$ \slash\ $\checkmark$)    & $\checkmark$ \\ \hline
    \ce{CoSb3}          & 16 & $Im\bar3$    & I & $\checkmark$ ($\checkmark$ \slash\ $\checkmark$)    & $\checkmark$ \\ \hline
    \ce{CsPbI3}         & 20 & $Pnma$       & I & $\times$ ($\times$ \slash\ $\checkmark$)            & $\times$ \\ \hhline{======}
    \ce{MnAl}           & 2 & $P4/mmm$      & II & $\checkmark$ ($\checkmark$ \slash\ $\checkmark$)   & $\checkmark$ \\ \hline
    \ce{HoHSe}          & 3 & $P\bar6m2$    & II & $\checkmark$ ($\checkmark$ \slash\ $\checkmark$)   & $\checkmark$ \\ \hline
    \ce{ErCdRh2}        & 4 & $Fm\bar3m$    & II & $\checkmark$ ($\checkmark$ \slash\ $\checkmark$)   & $\checkmark$ \\ \hline
    \ce{Eu2MgTl}        & 4 & $Fm\bar3m$    & II & $\checkmark$ ($\checkmark$ \slash\ $\checkmark$)   & $\checkmark$ \\ \hline
    \ce{Pm2NiIr}        & 4 & $Fm\bar3m$    & II & $\checkmark$ ($\checkmark$ \slash\ $\checkmark$)   & $\checkmark$ \\ \hline
    \ce{LaAl3Au}        & 5 & $I4mm$        & II & $\checkmark$ ($\checkmark$ \slash\ $\checkmark$)   & $\checkmark$ \\ \hline
    \ce{Ca3Ni7B2}       & 12 & $R\bar3m$    & II & $\checkmark$ ($\checkmark$ \slash\ $\checkmark$)   & $\checkmark$ \\ \hline
    \ce{LaSiIr}         & 12 & $P2_13$      & II & $\checkmark$ ($\checkmark$ \slash\ $\checkmark$)   & $\checkmark$ \\ \hline
    \ce{SmVO4}          & 12 & $I4_1/amd$   & II & $\checkmark$ ($\checkmark$ \slash\ $\checkmark$)   & $\checkmark$ \\ \hline
    \ce{Zr4O}           & 15 & $R\bar3$     & II & $\times$ ($\times$ \slash\ $\times$)               & $\times$     \\ \hline
    \ce{Ba3Ta2NiO9}     & 15 & $P\bar3m1$   & II & $\checkmark$ ($\checkmark$ \slash\ $\checkmark$)   & $\checkmark$     \\ \hline
    \ce{LiSm2IrO6}      & 20 & $P2_1/c$     & II & $\times$ ($\checkmark$ \slash\ $\checkmark$)       & $\checkmark$ \\ \hline
    \wcline{1-6}
\multicolumn{1}{r}{\textbf{Overall}}        & & & & $21 / 25 = 84.0\%$ & $23 / 25 = 92.0\%$ \\

\bottomrule
\end{tabular}

\label{tab:uspex_results}
\end{table}

\clearpage

\section*{Visualization of solved structures}
\addcontentsline{toc}{section}{Visualization of solved structures}

\begin{figure}[!ht]%
\centering
\includegraphics[width=0.90\linewidth]{suppl/structures_review/1.pdf}
\caption[]{}
\end{figure}%
\begin{figure}[ht]\ContinuedFloat
\centering
\includegraphics[width=0.90\linewidth]{suppl/structures_review/2.pdf}
\end{figure}%\
\begin{figure}[ht]\ContinuedFloat
\centering
\includegraphics[width=0.90\linewidth]{suppl/structures_review/3.pdf}
\end{figure}%
\begin{figure}[ht]\ContinuedFloat
\centering
\includegraphics[width=0.90\linewidth]{suppl/structures_review/4.pdf}
\end{figure}%
\begin{figure}[ht]\ContinuedFloat
\centering
\includegraphics[width=0.90\linewidth]{suppl/structures_review/5.pdf}
\end{figure}%
\begin{figure}[ht]\ContinuedFloat
\centering
\includegraphics[width=0.90\linewidth]{suppl/structures_review/6.pdf}
\end{figure}%
\begin{figure}[ht]\ContinuedFloat
\centering
\includegraphics[width=0.90\linewidth]{suppl/structures_review/7.pdf}
\end{figure}%
\begin{figure}[ht]\ContinuedFloat
\centering
\includegraphics[width=0.90\linewidth]{suppl/structures_review/8.pdf}
\end{figure}%
\begin{figure}[ht]\ContinuedFloat
\centering
\includegraphics[width=0.90\linewidth]{suppl/structures_review/9.pdf}
\end{figure}%
\begin{figure}[ht]\ContinuedFloat
\centering
\includegraphics[width=0.90\linewidth]{suppl/structures_review/10.pdf}
\end{figure}%
\begin{figure}[ht]\ContinuedFloat
\centering
\includegraphics[width=0.90\linewidth]{suppl/structures_review/11.pdf}
\end{figure}%
\begin{figure}[ht]\ContinuedFloat
\centering
\includegraphics[width=0.90\linewidth]{suppl/structures_review/12.pdf}
\end{figure}%
\begin{figure}[ht]\ContinuedFloat
\centering
\includegraphics[width=0.90\linewidth]{suppl/structures_review/13.pdf}
\end{figure}%
\begin{figure}[ht]\ContinuedFloat
\centering
\includegraphics[width=0.90\linewidth]{suppl/structures_review/14.pdf}
\end{figure}%
\begin{figure}[ht!]\ContinuedFloat
\centering
\includegraphics[width=0.90\linewidth]{suppl/structures_review/15.pdf}
\caption{All 120 stable structures in Datasets I, II, and III, along with their predicted structures using the element-substitution- and Wyckoff-position-generator-based CSP algorithms (depicted with VESTA \cite{momma2011VESTAThreedimensional} version 3.5.8). For each prediction algorithm, the structures with the two lowest DFT energies are shown.}
\label{fig:S4_structure}
\end{figure}

\clearpage
% \FloatBarrier

%% references
% \printbibliography
% \bibliography{refs}